\documentclass[aps,pre,10pt,showpacs,floatfix,twocolumn,longbibliography,nofootinbib]{revtex4-1}
\usepackage{amsmath}
\usepackage{amsfonts}
\usepackage{mathrsfs}
\usepackage{graphicx,color}
\usepackage{bm}
\usepackage{verbatim}
\usepackage{pstricks}

\newcommand{\vctr}[1]{\bm{#1}}

\newcommand{\diver}{\operatorname{div}}
\newcommand{\curl}{\operatorname{curl}}

\newcommand{\cov}{\operatorname{D}\!}
\newcommand{\dd}{\operatorname{d}\!}

\newcommand{\sgrad}{\nabla_\mathrm{{\!s}}}
\newcommand{\sdiv}{\diver_\mathrm{{\!s}}\!}
\newcommand{\scurl}{\curl_\mathrm{{s}}\!}
\newcommand{\n}{\bm{n}}
\newcommand{\nzero}{\bm{n}^{(0)}}
\newcommand{\e}{\bm{e}}
\newcommand{\rv}{\bm{r}}
\newcommand{\uv}{\bm{u}}
\newcommand{\tangent}{\bm{t}}
\newcommand{\binormal}{\bm{b}}
\newcommand{\trans}{^\mathsf{T}}
\newcommand{\mindis}{\sgrad\n^{(0)}}
\newcommand{\N}{\mathbf{N}}
\newcommand{\sgradn}{\sgrad\n}

\newcommand{\surface}{\mathscr{S}}
\newcommand{\surfaceC}{\mathscr{S}_{\curve}}
\newcommand{\curve}{\mathscr{C}}
\newcommand{\spin}{\bm{\omega}}
\newcommand{\conn}{\bm{w}}

\newcommand{\normal}{\bm{\nu}}
\newcommand{\ctens}{\sgrad\normal}
\newcommand{\bzero}{\bm{0}}

\begin{document}

\title{Bistable curvature potential at hyperbolic points of nematic shells}
\author{Andr\'e M. \surname{Sonnet}}
\email[e-mail: ]{Andre.Sonnet@strath.ac.uk}
\affiliation{Department of Mathematics and Statistics, University of Strathclyde,
Livingstone Tower, 26 Richmond Street, Glasgow G1 1XH, Scotland}
\author{Epifanio G. \surname{Virga}}
\email[e-mail: ]{eg.virga@unipv.it}
\thanks{On leave from Dipartimento di Matematica, Universit\`a di Pavia, Via Ferrata 5, I-27100 Pavia, Italy.}
\affiliation{Mathematical Institute, University of Oxford, Oxford, UK}

\date{\today}

\begin{abstract}
Nematic shells are colloidal particles coated with nematic liquid
crystal molecules which may freely glide and rotate on the
colloid's surface while keeping their long axis on the local
tangent plane.
We describe the nematic order on a shell by a unit director field
on an orientable surface. Equilibrium fields can then be
found by minimising the elastic energy, which in general is a function of the surface
gradient of the director field.
We learn how to extract systematically out of this energy a \emph{fossil} component,
related only to the surface and its curvatures, which expresses a \emph{curvature potential}
for the molecular torque. At hyperbolic points on the colloid's surface,
and only there, the alignment preferred by the curvature potential may fail to be
a direction of principal curvature. There the fossil energy becomes bistable.
\end{abstract}

\maketitle

\section{Introduction}\label{sec:introduction}
Nematic shells are (mostly undeformable) colloidal particles coated with a thin film of a nematic liquid crystal,
whose rod-like molecules are subject to a \emph{degenerate planar} anchoring exerted by the solid substrate.
This anchoring allows the molecules to rotate freely on the local tangent plane, but makes
it hard for them to flip out of that plane. 

We still do not know whether the futuristic dream of the all-optical computer will ever come true.
For sure, Ozin's ``colloidal photonic crystal metropolis'' \cite{arsenault:towards} is a natural
technological evolution of Nelson's ``colloidal chemistry'' \cite{nelson:toward},
which envisioned replacing atoms with nematic decorated colloids.

This may explain the recent surge of experimental studies on nematic shells and their aim to produce different stable configurations with a \emph{tunable} number of \emph{defects}, which in Nelson's paradigm would replace the atomic valence.\footnote{Nelson's original contribution \cite{nelson:toward} showed that a spherical nematic shell can become a \emph{tetravalent} colloid. A similar construction is valid for the wetting layer that surrounds spherical colloids dispersed in a liquid crystal at a temperature above the nematic-isotropic phase transition \cite{huber:tetravalent}.} See, for example, \cite{fernandez-nieves:novel,lopez-leon:frustrated,lopez-leon:nematic-smectic,liang:nematic-smectic,sec:defect,lopez-leon:defect,noh:taming,noh:influence,koning:spherical}, only to mention a few representative experimental works.

A number of simulation studies are also available, which bridge experiment and theory \cite{skacej:controlling,shin:topological,bates:defects,dhakal:nematic,li:topological,li:defect-free,mbanga:simulating,wand:monte}.

Apart from the director elastic theories that will be discussed more closely  below, as they form the background of our contribution, theories from different lineages have also been proposed; they can be broadly classified in two groups: the group of phenomenological theories formulated in terms of an appropriate two-dimensional version of de Gennes' order tensor, especially suitable to the description of defects and their controllability \cite{kralj2011curvature,napoli2013curvature,jesenek:defect,mesarec:effective}, and the group of molecular theories phrased within Onsager's excluded volume density functional formalism, duly extended to surfaces, which are also interesting for the light they shed  on the interplay between defects and density depletion \cite{zhang:onsager,zhang:solution,liang:rigid,ye:nematic}.

Luckily, there are excellent reviews that might safely navigate the reader into the vast literature on nematic shells \cite{lopez-leon:drops,lagerwall:new,mirantsev:defect,serra2016curvature,urbanski2017liquid}.

Our contribution in this paper is placed within the director elastic theories of nematic shells,
at the heart of the debate on what should be the most appropriate form of the energy.
Two groups of theories have been proposed so far: in one group, pioneered by \cite{nelson:fluctuations},
the only molecular distortions that matter are measured within the surface's metric;
in the other group, pioneered by \cite{helfrich:intrinsic},  the way molecules couple
with the curvature of the surface should also be accounted for in the energy balance.
Theories in the former group are conventionally called \emph{intrinsic}, whereas theories
in the latter group are called \emph{extrinsic}. An interesting third way was opened by Selinger
and his co-workers \cite{nguyen:nematic}, who offered reasons to subject to
\emph{unequal} weights these different sources of energy. Much in the spirit of this latter view,
here we start from an elementary, direct way to identify one  component of these energies,
which we shall call the \emph{fossil} energy because it is inextricably interwoven with the
background surface. Not surprisingly perhaps, we find in an appropriate rendition of
Levi-Civita's parallel transport the elementary geometric tool that extracts the
fossil energy out of any proposed phenomenological energy.

This tool is more general than the application of it we make to the analogue of Frank's elastic energy.
When written in the frame of principal directions of curvatures,
the fossil energy reveals itself as a \emph{curvature potential} for the torque acting
on molecules and responsible for their fossil preferred alignment.
The clarity that may be gained with this approach has an interesting consequence:
we predict that at hyperbolic points of the shell, and only there,
the fossil preferred alignment of molecules may fail to be a principal direction of curvature. By symmetry, whenever this applies, the fossil energy becomes \emph{bistable}.
The preferred alignments 
depend on the elastic anisotropy, possibly only the anisotropy of the fossil energy,
in the dissociated approach of \cite{nguyen:nematic}.

The structure of the paper is the following. In Sec.~\ref{sec:ground_state},
we present our elementary way to construct the fossil energy, starting from the notion
of parallel transport, which is recalled in a simple, but rigorous language.
A form of elastic energy is considered in Sec.~\ref{sec:energies}, where we derive
formally the associated  fossil energy and represent it so as to make explicit its
nature of curvature potential for the molecular torque. In Sec.~\ref{sec:curvature_potential},
we study the general equilibria for the curvature potential; their multiplicity and stability
is analysed at hyperbolic points of a shell, where these properties become more intriguing.
Sec.~\ref{sec:conclusions} draws the conclusions of this paper.
There we also discuss in some detail where our study stands in the context of many other similar ones.
We also outline problems that in our view could profit from the insights afforded by our approach.
The paper is closed by an appendix, somewhat more technical in character.  

\section{Parallel transport and local ground state}\label{sec:ground_state}
In three-dimensional space, the \emph{ground state} of a nematic liquid crystal, that is, the configuration for the director $\n$ that realises the minimum distortion energy, is easily identified: this is any configuration in which $\n$ is uniform in space, no matter in which orientation, and so $\nabla\n\equiv\mathbf{0}$. As is well known from the classical theory of Frank (see, for example \cite[Chap.~3]{virga:variational}), the distortional cost associated with the ground state in three dimensional space is \emph{everywhere} zero.

On a nematic shell, here represented by a smooth surface $\surface$ with outer unit normal $\normal$,
on which molecules are constrained to lie subject to the only constraint that
\begin{equation}\label{eq:orthogonality_constraint}
\n\cdot\normal\equiv0,
\end{equation}
the ground state is not identified with equal ease. But it is even more important to have a clear concept of
the ground state,
as we shall see that it brings along a distortional cost, not uniformly distributed on $\surface$,
reminiscent of a \emph{fossil} energy, inextricable from its geometric background. 

We start with the nearly obvious remark that $\sgrad\n\equiv\bzero$, where
$\sgrad$ denotes the surface gradient,
cannot be the appropriate definition for the ground state of nematic shells,
despite the fact that $\sgrad\n$ appropriately measures the degree of
distortion of molecules interacting in three-dimensional space.
The reason for this follows at once from \eqref{eq:orthogonality_constraint},
under the assumption that both $\n$ and $\normal$ are smooth unit vector
fields on $\surface$. Differentiating both sides of \eqref{eq:orthogonality_constraint}, we see that
\begin{eqnarray}\label{eq:why_not}
(\sgrad\n)\trans\normal+(\sgrad\normal)\n\equiv\bzero,
\end{eqnarray}
where $\ctens$ is the (symmetric) curvature tensor of $\surface$.
A local distortion state with $\sgrad\n=\bzero$ would also require $(\ctens)\n=\bzero$,
which in turns wants $\n$ to be oriented along a direction of zero curvature of $\surface$,
which may in general fail to exist. On the other hand, at the points on $\surface$ where $\sgradn\neq\bzero$,
the two-dimensional layer covering $\surface$ carries a distortional energy, if seen from the embedding three-dimensional space, because there the molecules with average orientation  $\n$ necessarily fail to be parallel to one another, disagreeing  with the three-dimensional ground state (as there cannot be any director field $\n^\ast$ in space that agrees with $\n$ on $\surface$ and has $\nabla\n^\ast=\bzero$).
 
Thus, we are presented with a clear task: given $\n$ at a point on $\surface$, identify geometrically the least distortion $\mindis$ that can locally be compatible with this assignment (and may well fail to be a gradient). We shall see shortly below that a natural geometric tool to accomplish this task is provided by Levi-Civita's parallel transport. The energy associated with $\mindis$, which by necessity incorporates appropriate local measures of the curvature of $\surface$, will be called the \emph{fossil} energy. This is the energy that according to our theory is stored in the \emph{local} ground state, which may be different at different points on $\surface$.  

\subsection{Parallel transport}\label{sec:parallel_transport}
Little new, if anything, can be said about a subject that has been extensively studied for over a century.
Already Cartan in his memoir \cite{cartan:geometrie} gave a lucid presentation of the role played by
Levi-Civita's notion of parallelism in Riemannian geometry \cite{levi-civita:nozione}; there the distinction
between what are now often called intrinsic and extrinsic properties is admirably illuminated (though in a different,
more descriptive language).
Here we follow closely the kinematic interpretation given in \cite{rosso2012parallel} to Levi-Civita's
parallel transport on surfaces. We build on much earlier work of Persico~\cite{persico:realizzazione},
lately also revived in \cite{pfister:spatial}.

Let two perpendicular unit vectors, $\uv_1$ and $\uv_2:=\normal\times\uv_1$,
tangent to $\surface$ be prescribed along a smooth curve $\curve$ through a given point $p\in\surface$.
The triad $(\uv_1,\uv_2,\normal)$ constitutes a \emph{movable} frame along $\curve$ constrained to have
an axis aligned with the surface's normal. The movable frame is parametrized in the arc length $s$ of $\curve$
and differentiation with respect to $s$ is denoted by a prime $'$.
Letting $\tangent$ be the unit tangent to $\curve$, $\normal'$ is delivered by
\begin{equation}\label{eq:nu_prime}
\normal'=(\ctens)\tangent,
\end{equation}
but both $\uv_1'$ and $\uv_2'$ have an extra degree of freedom. Elementary kinematics shows that
\begin{equation}\label{eq:frame_evolution}
\uv_1'=\spin\times\uv_1,\quad\uv_2'=\spin\times\uv_2,\quad\normal'=\spin\times\normal,
\end{equation}
where $\spin$ is the \emph{spin vector} defined along $\curve$. Only the component $\spin_\parallel$ of the spin vector tangent to $\surface$ is prescribed by \eqref{eq:nu_prime} and the third of \eqref{eq:frame_evolution}, whereas the component parallel to $\normal$ remains completely free. Different choices of $\spin\cdot\normal$ impart different spins to the evolution of the movable frame along $\curve$. Clearly, the evolution corresponding to the least admissible spin is obtained for $\spin\cdot\normal=0$, which is said to characterize the \emph{parallel transport} of the frame. The least spin vector $\spin_\parallel$ is easily obtained from \eqref{eq:frame_evolution} as
\begin{equation}\label{eq:spin_parallel}
\spin_\parallel=\normal\times\normal'.
\end{equation}
Combining \eqref{eq:nu_prime} and \eqref{eq:spin_parallel}, we easily arrive at
\begin{equation}\label{eq:spin_parallel_final}
\spin_\parallel=\N(\ctens)\tangent,
\end{equation}
where $\N$ is the skew-symmetric tensor associated with the normal $\normal$, so that 
\begin{equation}\label{eq:skew_normal}
\N\bm{v}=\normal\times\bm{v}, 
\end{equation}
for all vectors $\bm{v}$.
More generally, a vector $\uv$ tangent to $\surface$ at $p$ is said to be parallel transported along $\curve$ if it is rigidly conveyed by a parallel transported frame, which amounts to prescribe for it the following evolution law,
\begin{equation}\label{eq:parallel_trasport_evolution}
\uv'=\spin_\parallel\times\uv.
\end{equation}	

Let now $\n$ be prescribed at $p$. We want to construct a distortion field $\mindis$, if it exists,
that delivers in the vicinity of $p$ on $\surface$ the same vector delivered by parallel transporting
$\n$ along all possible curves emanating from $p$. This requires that the following equation
\begin{equation}\label{eq:minimu_distortion_equation}
(\mindis)\tangent=\spin_\parallel\times\n=-\n\times\N(\ctens)\tangent
\end{equation}
be valid for all unit tangent vectors $\tangent$. Since, by \eqref{eq:skew_normal}, 
\begin{equation}\label{eq:minimum_distortion_computation}
\n\times\N\bm{v}=(\normal\otimes\n)\bm{v},
\end{equation}
for all vectors $\bm{v}$, \eqref{eq:minimu_distortion_equation} is satisfied by
\begin{equation}\label{eq:minimum_distortion_solution}
\mindis=-\normal\otimes(\ctens)\n.
\end{equation}	
This is the \emph{minimal} distortion induced locally by the curvature of $\surface$ that would agree
with the orientation $\n$ prescribed at a given point $p\in\surface$. The reader
should not be misled by our (perhaps infelicitous) notation for the measure of minimal distortion,
as in general this is \emph{not} the surface gradient of a field.
As recalled in Appendix~\ref{sec:app_developable}, for this to be the case,
the surface $\surface$ must possess zero Gaussian curvature $K$, that is,
it must be \emph{developable}. For a non-developable surface,
$\mindis$ in \eqref{eq:minimum_distortion_solution} merely represents a measure of local distortion,
which cannot be globally integrated.

Suppose that $\sgradn$ is a \emph{genuine} measure of distortion, that is, the surface gradient of a local field $\n$, not necessarily in the form \eqref{eq:minimum_distortion_solution}. We define
\begin{equation}\label{eq:covariant_derivative_definition}
\cov\n:=\sgradn-\mindis=\sgradn+\normal\otimes(\ctens)\n
\end{equation}
and attribute to it the meaning of an \emph{extra} distortion, one that comes on top of the minimal distortion compatible with the value assigned to the field $\n$ at $p\in\surface$. $\cov\n$ is better known as the \emph{covariant} gradient of $\n$ on $\surface$, as it can be given an \emph{intrinsic} representation, independent of the way the surface is embedded in space.

To see this, we consider a frame $(\uv_1,\uv_2,\normal)$ that is locally parallel transported on $\surface$. In this frame we represent $\n$ as
\begin{equation}\label{eq:n_alpha_representation}
\n=\cos\alpha\,\uv_1+\sin\alpha\,\uv_2,
\end{equation}
so that 
\begin{equation}\label{eq:grad_n_alpha_representation}
\sgradn=\n_\perp\otimes\sgrad\alpha+\cos\alpha\sgrad\uv_1+\sin\alpha\sgrad\uv_2,
\end{equation}
where $\n_\perp:=\N\n$. By applying \eqref{eq:minimum_distortion_solution} to both $\uv_1$ and $\uv_2$, we easily rewrite \eqref{eq:grad_n_alpha_representation} as 
\begin{equation}\label{eq:grad_n_alpha_rewritten}
\sgradn=\n_\perp\!\otimes\sgrad\alpha-\normal\otimes(\ctens)\n,
\end{equation}
which combined with \eqref{eq:covariant_derivative_definition} shows that 
\begin{equation}\label{eq:covariant_derivative_representation}
\cov\n=\n_\perp\!\otimes\sgrad\alpha.
\end{equation}
The latter two equations have an interesting consequence: they show that $\cov\n$ and $\mindis$ decompose $\sgradn$
into two orthogonal components. Thus, $\cov\n$ can alternatively be obtained by projecting $\sgradn$ on
the plane tangent to $\surface$,
\begin{equation}\label{eq:covariant_derivative_projection}
\cov\n=(\mathbf{I}-\normal\otimes\normal)\sgradn.
\end{equation}

Assuming that the  distortion energy density $W$ is arising from the way molecules interact (and, as it were, ``see'' each other)
in the ambient three-dimensional space, an assumption corroborated by the spin interaction model studied in \cite{selinger:monte},
we are led to consider $W$ as a function of the form $W(\n,\sgradn)$. The least distortion in \eqref{eq:minimum_distortion_solution}
then identifies the \emph{local} ground state of the system, solely dictated by the orientation of $\n$ on $\surface$ and the
curvature tensor of $\surface$.
Computing $W$ on the local ground state we identify the \emph{fossil} energy density, an energy that 
can be attributed to the curvature of the shell and cannot be extracted from it, as long as $\surface$
is thought of as undeformable. In general, the fossil energy is given by
\begin{equation}\label{eq:fossil_energy_definition}
W_0(\n,\ctens):=W(\n,\mindis).
\end{equation}
We shall see in the following section that for the most general $W$ quadratic in $\sgradn$ the orthogonal decomposition in \eqref{eq:grad_n_alpha_rewritten}, appropriately extended, will be reflected in splitting $W$ into the sum of $W_0$ and an extra energy depending only on the intrinsic measure of distortion.

\section{Energetics}\label{sec:energies}
We follow \cite{chen:nematic,napoli2012surface} to obtain the two-dimensional energy density $W$ by
a dimension reduction of Frank's classical three-dimensional formula, viewing the shell as a
\emph{thin} layer coating $\surface$. Denoting by $K_i$, $i=1,\dots,4$, Frank's elastic constants
(see Chap.~3 of \cite{virga:variational}), and by $h$ the layer's thickness (assumed to be much
smaller than the smallest radius of curvature of $\surface$), we easily arrive at 
\begin{equation}\label{eq:W_dimension_reduction}
\begin{split}
W(\n,\sgradn)&=
\frac{1}{2}k_1    (\sdiv\vctr{n})^2 +
\frac{1}{2}k_2    (\vctr{n}\cdot\scurl\vctr{n})^2\\
&+
\frac{1}{2}k_3    (\vctr{n}\times\scurl\vctr{n})^2,
\end{split} 
\end{equation}
where $k_i=K_ih$, $i=1,2,3$, and no saddle-splay energy is inherited from the parent three-dimensional
energy (see \cite{napoli2012surface} for further details). By Ericksen's inequalities for the constants
$K_i$ \cite{ericksen:inequalities}, all three reduced elastic constants in \eqref{eq:W_dimension_reduction}
must be non-negative, $k_i\geq0$. $W$ represents the elastic energy per unit area of $\surface$, and so all
the $k_i$ have the physical dimensions of an energy (whereas all the $K_i$ have the physical dimensions of a force). 

\subsection{Fossil energy and curvature potential}\label{sec:fossil_energy}
Here, primarily, we want  to isolate the fossil energy part of $W$. 
To this end, we first need to generalize the gradient decomposition in \eqref{eq:grad_n_alpha_rewritten}
to the case where the movable frame $(\uv_1,\uv_2,\normal)$ used to parametrize $\n$ in \eqref{eq:n_alpha_representation}
is not locally parallel transported.

This is achieved in a standard way by use of a \emph{spin connection} $\conn$, defined as in \cite{kamien:geometry}
by\footnote{The spin connection $\conn$ is nothing but the \emph{geometric vector potential} introduced by
Nelson and Peliti~\cite{nelson:fluctuations}; see also \cite{vitelli:defect} and \cite{nguyen:nematic}
for the use of this latter name for $\conn$.}
\begin{equation}\label{eq:spin_connector_definition}
\conn:=(\sgrad\uv_2)\trans\uv_1=-(\sgrad\uv_1)\trans\uv_2.
\end{equation}
It follows from using the orthonormality conditions linking the vectors of the movable frame that 
\begin{subequations}\label{eq:frame_gradients}
\begin{align}
\sgrad\uv_1&=-\uv_2\otimes\conn-\normal\otimes(\ctens)\uv_1,\\
\sgrad\uv_2&=\uv_1\otimes\conn-\normal\otimes(\ctens)\uv_2,
\end{align}
\end{subequations}	
which clearly show the mismatch with the law of parallel transport \eqref{eq:minimum_distortion_solution}.
Combining \eqref{eq:frame_gradients} and \eqref{eq:grad_n_alpha_representation}, we extend \eqref{eq:grad_n_alpha_rewritten} as follows,
\begin{equation}
\sgradn=\n_\perp\otimes(\sgrad\alpha-\conn)-\normal\otimes(\ctens)\n,
\end{equation}
from which we readily arrive at 
\begin{subequations}\label{eq:div_curl_connector}
\begin{gather}
\sdiv\n=\n_\perp\cdot(\sgrad\alpha-\conn),\\
\scurl\n=(\sgrad\alpha-\conn)\times\n_\perp+\normal\times(\ctens)\n.
\end{gather}
\end{subequations}
From the latter equation, which for a unit vector field $\n$ tangent to $\surface$ is an
orthogonal decomposition of $\scurl\n$, it easily follows that 
\begin{subequations}\label{eq:curl_computations}
\begin{gather}
|\n\cdot\scurl\n|=|(\ctens)\n\times\n|,\\
|\n\times\scurl\n|^2=[(\sgrad\alpha-\conn)\cdot\n]^2+(\n\cdot(\ctens)\n)^2.
\end{gather}
\end{subequations}
Use of \eqref{eq:curl_computations} and 
\begin{equation}
(\n\cdot(\ctens)\n)^2=|(\ctens)\n|^2-|(\ctens)\n\times\n|^2
\end{equation}
in \eqref{eq:W_dimension_reduction} delivers the following  general representation formula for $W$,
\begin{align}\label{eq:W_general_representation_formula}
W&=\frac12k_1[\n_\perp\cdot(\sgrad\alpha-\conn)]^2+\frac12(k_2-k_3)|(\ctens)\n\times\n|^2\nonumber\\&+\frac12k_3\{[\n\cdot(\sgrad\alpha-\conn)]^2+|(\ctens)\n|^2\},
\end{align}
which neatly reveals the \emph{fossil} energy as 
\begin{equation}\label{eq:fossil_energy_revealed}
\begin{split}
W_0(\n,\ctens)&
=\frac12(k_2-k_3)|(\ctens)\n\times\n|^2\\&
+\frac12k_3|(\ctens)\n|^2.
\end{split}
\end{equation}
For definiteness, we shall call the difference 
\begin{equation}\label{eq:distortional_energy}
W-W_0=\frac12k_1[\n_\perp\cdot(\sgrad\alpha-\conn)]^2+\frac12k_3[\n\cdot(\sgrad\alpha-\conn)]^2
\end{equation}
the \emph{distortional} energy.

As already remarked in \cite{napoli2012surface}, in contrast with the planar case, a twist contribution arises if $\surface$ is generally curved. However, in our theory such a twist energy is always a part of the fossil energy. Moreover, and more importantly, as clearly indicated by \eqref{eq:fossil_energy_revealed}, for $k_2\geq k_3$ the fossil energy $W_0$ is minimized when $\n$ is oriented along the eigenvector of $\ctens$ with the least absolute eigenvalue, but for $k_2<k_3$ this is no longer the trivial minimizing choice, and new interesting scenarios may arise, as will be shown below.

Having established \eqref{eq:W_general_representation_formula} for a general spin connection $\conn$, we can now afford representing $\n$ in a local frame more geometrically telling, which is not necessarily parallel transported on the surface. This is the frame $(\e_1,\e_2,\normal)$, where $(\e_1,\e_2)$ are the principal directions of curvature, with $(\kappa_1,\kappa_2)$ the 
corresponding principal curvatures. Without loss of generality, we can
assume that $\kappa_1^2\le\kappa_2^2$. So
\begin{subequations}\label{eq:n_parametrization}
\begin{align}
\sgrad\vctr{\nu}&=\kappa_1\vctr{e}_1\otimes\vctr{e}_1+\kappa_2\vctr{e}_2\otimes\vctr{e}_2,\\
\vctr{\nu}&=\vctr{e}_1\times\vctr{e}_2,\\
\vctr{n}&=\cos\theta\vctr{e}_1+\sin\theta\vctr{e}_2,
\end{align}
\end{subequations}
where $\theta$ is the angle that the director $\vctr{n}$ makes with
$\vctr{e}_1$. Making use of \eqref{eq:n_parametrization} in \eqref{eq:fossil_energy_revealed}, we thus convert the fossil energy density $W_0$ into a function of $\theta$, parametrized in the principal curvatures,
\begin{equation}\label{eq:curvature_potential}
\begin{split}
W_0&=\frac18(k_2-k_3)(\kappa_1-\kappa_2)^2\sin^2 2\theta\\
&+\frac{1}{4}k_3\left[
\kappa_1^2+\kappa_2^2+(\kappa_1^2-\kappa_2^2)\cos 2\theta
\right].
\end{split}
\end{equation}
A representation equivalent to \eqref{eq:curvature_potential} was also obtained in \cite{segatti:equilibrium} (see their equation (5), then reiterated in (6.19) of \cite{segatti2016analysis}), though not endowed with the same meaning.

In this form, $W_0$ appears like the potential energy for a torque on the nematic director, a torque that is not imparted by external agents,
such as a field, but which
stems
from the curvature of the shell. Having this interpretation in mind,
we shall refer to $W_0$ in \eqref{eq:curvature_potential} as the \emph{curvature potential}.
Its equilibria are the orientations that make the local fossil energy achieve a critical value
(while the corresponding torque vanishes); its minimizer is the orientation that the local
fossil energy would enforce, \emph{if} it were the only energy at play.

It was first shown in \cite{napoli2012extrinsic} how the fossil energy may be minimized when $\n$ is aligned
along the direction of minimum absolute principal curvature ($\e_1$, in our parametrization). 
We shall see  in the following section that this is not generally the case:
at hyperbolic points of a shell, the curvature potential can induce a preferred direction
coincident with neither of the principal directions of curvature.

A new interesting perspective on the energetics of \emph{flexible} nematic shells was recently
opened up by Selinger and his co-workers \cite{nguyen:nematic}. Building upon the decomposition of
$\sgradn$ obtained from \eqref{eq:covariant_derivative_definition},
it was proposed in \cite{nguyen:nematic} to write $W(\n,\sgradn)$ as the sum of two components,
weighted by \emph{different} elastic constants. One component, the \emph{intrinsic} energy $W(\n,\cov\n)$,
would be associated with the variations of $\n$ \emph{inside} the surface, whereas the other component,
our fossil energy $W(\n,\mindis)$, would be associated with the variations of $\n$, as it were,
\emph{outside} the surface. Such a dual treatment of the energy rests on the physical presumption
that different costs should assigned to distortions of  different origins. In our theory,
we have set equal the elastic constants of both energies, for simplicity. Adopting the point of
view of \cite{nguyen:nematic} would simply amount to divorce the elastic constants of the fossil
energy from those of the intrinsic energy, resulting in more freedom of action for the curvature
potential $W_0$ in \eqref{eq:curvature_potential}.  

\begin{figure}
\includegraphics[width=0.48\textwidth]{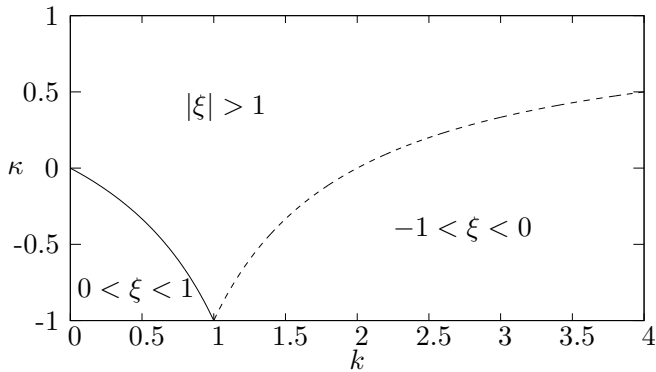}
\caption{\label{fig:solutionPlane}%
Value of $\xi$ defined by \eqref{eq:xi} as a function of
the ratio $k=k_2/k_3$ of the elastic constants and 
the ratio $\kappa=\kappa_1/\kappa_2$ of the principal
curvatures. In the lower part of the $k$-$\kappa$ plane
where $|\xi|\le 1$, a stationary director orientation exists
that is not along a principal direction of curvature.
This orientation is stable in the region under the solid
line where $0<\xi<1$, and it is unstable in the region
under the dashed line where $-1<\xi<0$.}
\end{figure}

\section{Curvature potential equilibria}\label{sec:curvature_potential}
The curvature potential \eqref{eq:curvature_potential} depends on the angle $\theta$ that the director makes with
the principal direction $\vctr{e}_1$ corresponding to the principal
curvature $\kappa_1$ with, as chosen above, $\kappa_1^2\le\kappa_2^2$.
To find the director orientations favoured by the curvature potential, we consider
\begin{equation}\label{eq:dW0dtheta}
\frac{dW_0}{d\theta}=
\frac{1}{2}\sin 2\theta\left[
k_3(\kappa_2^2-\kappa_1^2)+\cos 2\theta(k_2-k_3)(\kappa_1-\kappa_2)^2
\right]
\end{equation}
and 
\begin{equation}
\frac{d^2W_0}{d\theta^2}=
\cos 2\theta\,k_3(\kappa_2^2-\kappa_1^2)
+\cos 4\theta(k_2-k_3)(\kappa_1-\kappa_2)^2.
\end{equation}

\subsection{Equilibrium multiplicity and bifurcation}\label{sec:multiplicity_bifurcation}
We start by discussing \eqref{eq:dW0dtheta}. First of all,
at umbilical points, where $\kappa_1=\kappa_2$, the derivative
is identically zero and there is no preferred direction for
the director.

Let us now assume that $\kappa_1\ne\kappa_2$. In the case of 
equal elastic constants, $k_2=k_3$, the only zeros
of the derivative are found when $\sin 2\theta=0$,
that is, at $0$, $\pi$, and $\pm\pi/2$, which means
that the director prefers to align along one of the
principal directions. The second derivative is positive
if $\theta$ is zero or $\pi$ and negative when $\theta$
is $\pm\pi/2$.
So the director aligns along the principal direction
corresponding to the  principal curvature with smaller
absolute value. The fossil energy densities for orientation
along the lines of principal curvature are
$\left.W_0\right|_{\theta=0}=\frac12 k_3 \kappa_1^2$
and 
$\left.W_0\right|_{\theta=\pi/2}=\frac12 k_3 \kappa_2^2$.

In the case of unequal elastic constants,
$k_2-k_3\ne 0$, there is the possibility that the term in
square brackets on the right-hand side of \eqref{eq:dW0dtheta}
might vanish. This is the case if
\begin{equation}\label{eq:cos2theta0}
\cos 2\theta_0=\xi
\end{equation}
where
\begin{equation}\label{eq:xi}
\xi:= \frac{k_3}{k_3-k_2}\times\frac{\kappa_2+\kappa_1}{\kappa_2-\kappa_1}. 
\end{equation}
Provided that $-1\le \xi\le 1$, there are critical
points at $\pm\theta_0$ and $\pi\pm\theta_0$.
The second derivative at the critical points then is
\begin{equation}
\frac{d^2W_0}{d\theta^2}=
(\kappa_2-\kappa_1)^2(k_2-k_3)\times\left\{ 
\begin{array}{ll}
1-\xi, &\theta=0,\\
1+\xi, &\theta=\pi/2,\\
\xi^2-1, &\theta= \pm\theta_0.
\end{array}\right.
\end{equation}
This shows that if $k_2>k_3$, then both lines of principal curvature
are minimisers. As before, the absolute minimum is attained for alignment
along 0 or $\pi$, but alignment along $\pm\pi/2$ now corresponds to
a local minimum. The directions corresponding to $\cos 2\theta_0=\xi$
yield the maximum of the fossil free energy density.

When $k_2<k_3$, the situation is reversed. Along the lines of 
principal curvature, maxima are attained, and the minimum free
energy density is found along the directions corresponding
to $\cos 2\theta_0=\xi$. The fossil energy density for orientation
along $\pm\theta_0$ is
\begin{equation}\label{eq:newEnergy}
\left.W_0\right|_{\theta=\pm\theta_0}= \frac{1}{2}k_3
\left[
\frac{1}{2}(\kappa_1^2+\kappa_2^2)+
\frac{1+\xi^2}{4\xi}(\kappa_1^2-\kappa_2^2)
\right].
\end{equation}

To discuss further the existence and stability of the solution
corresponding to \eqref{eq:cos2theta0}, we define $k:=k_2/k_3$
and $\kappa:=\kappa_1/\kappa_2$, so that we can
write
\begin{equation}\label{eq:xi_simple}
\xi=\frac{1}{1-k}\times\frac{1+\kappa}{1-\kappa}.
\end{equation}
Because both $k_2$ and $k_3$ are positive, we have
$k\in[0,\infty)$, and with our choice of $\kappa_2^2>\kappa_1^2$,
we have $\kappa\in[-1,1]$. The term
$\frac{1+\kappa}{1-\kappa}$ is always positive. It is less
than one if and only if $\kappa$ is negative. The
term $\frac{1}{1-k}$ is greater than one if $k<1$, that is
if $k_2<k_3$. It is negative if $k>1$, that is if $k_2>k_3$,
and its modulus can become arbitrarily small.

Figure \ref{fig:solutionPlane} shows the regions
in the $k$-$\kappa$-plane where this solution exists.
The solid line marks $\xi=1$ ($\theta_0=0$), and the dashed line
marks $\xi=-1$ ($\theta_0=\pi/2$).
The solution is stable for $k<1$ and unstable for
$k>1$. While the unstable solution exists for all admissible ratios
of the principal curvatures, if only $k$ is large enough, the stable
solution can only be found when $\kappa<0$, that is at hyperbolic
points on the surface.
When $\kappa=-1$, as long as $k\ne 1$, we
have $\xi=0$ ($\theta_0=\pi/4$).

It should be noted that in the region of the $k$-$\kappa$-plane delimited by the solid line in Fig.~\ref{fig:solutionPlane}, the fossil energy is minimised along \emph{two} directions, symmetrically oriented with respect to the principal direction of curvature $\e_1$ with minimum absolute principal curvature $\kappa_1$. The region $\surface_0$ on $\surface$ where this is the case, if existing, is delimited by the curve where
\begin{equation}\label{eq:bistable_region_delimiter}
\kappa=-\frac{k}{2-k},\quad k<1,
\end{equation} 
as easily follows from \eqref{eq:xi_simple}, upon setting $\xi=1$. $\surface_0$, if not empty, is the \emph{bistable} region for the fossil energy $W_0$.

\begin{figure}
\includegraphics[width=0.48\textwidth]{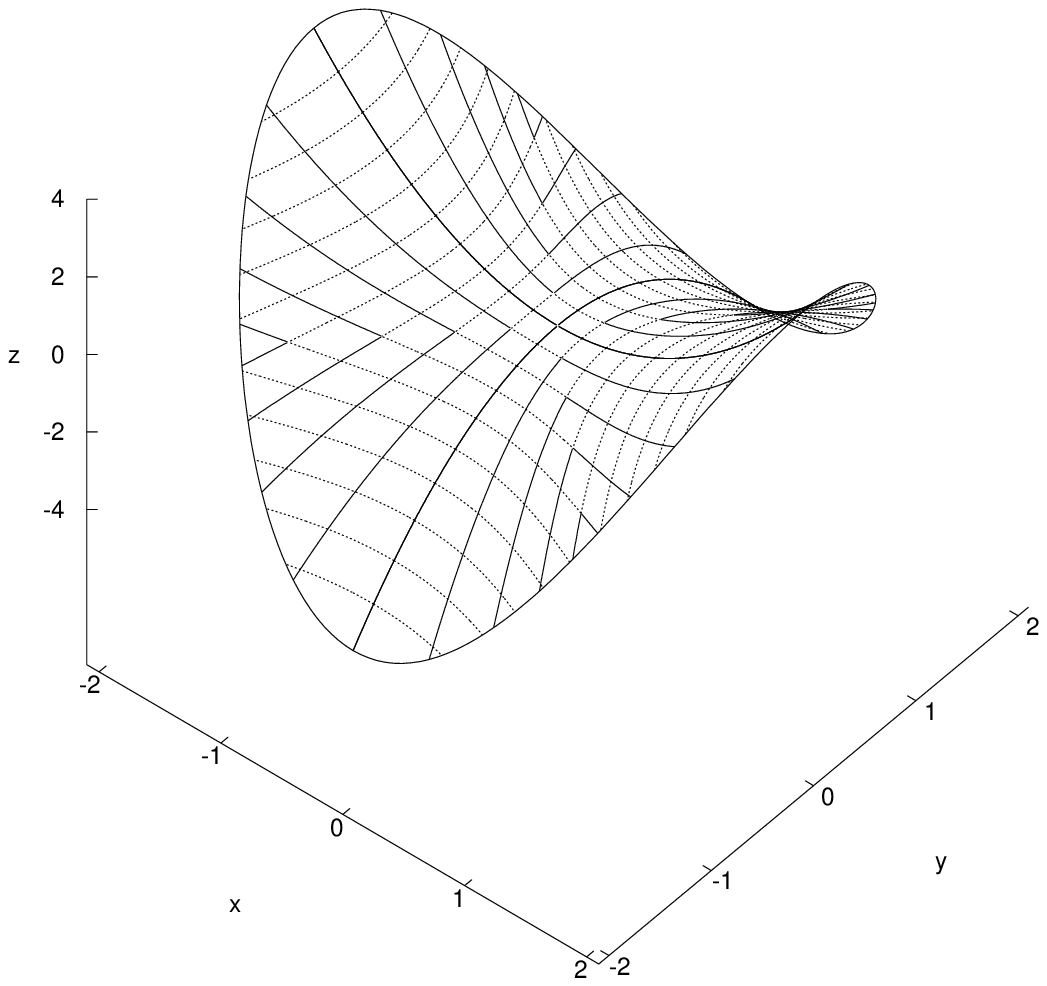}\\[10pt]
\includegraphics[width=0.48\textwidth]{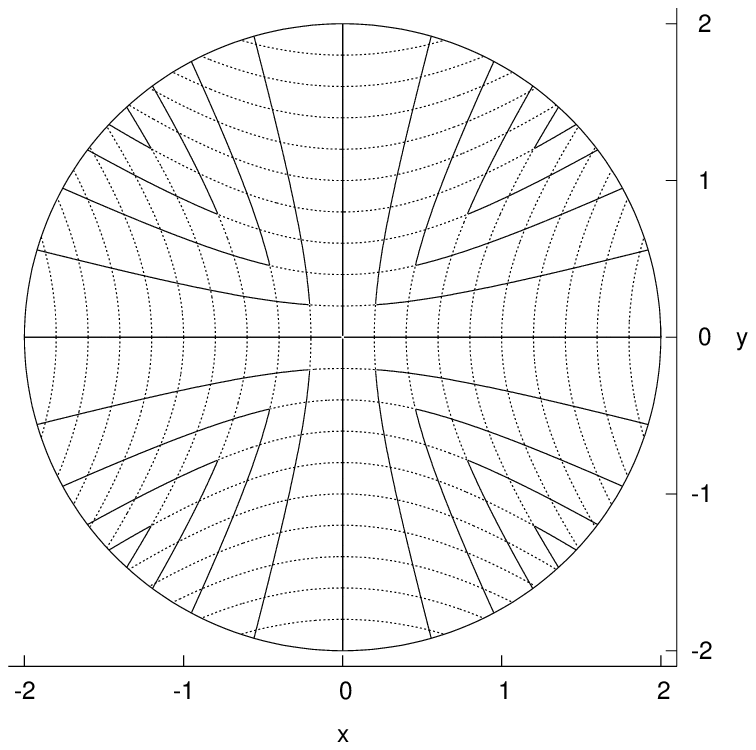}
\caption{\label{fig:principal}Principal lines of curvature on the hyperbolic paraboloid. 
Both panels of the figure show the same part of the manifold, the view in the second panel 
is along the $z$-axis, showing the projection onto the $x$-$y$-plane.
The solid lines correspond to the principal curvature with smaller absolute value. If $k_2=k_3$, the
curvature potential would always prompt the director to align along these lines.}
\end{figure}

\subsection{Curvature along lines of preferred directions}
Suppose that the director is at an angle $\theta_0$ corresponding
to \eqref{eq:cos2theta0},
\begin{equation}
\vctr{n}=\cos\theta_0\vctr{e}_1+\sin\theta_0\vctr{e}_2.
\end{equation}
The curvature $\kappa_n$ in this direction is
\begin{equation}
\kappa_n=\vctr{n}\cdot (\sgrad\vctr{\nu})\vctr{n}=
\kappa_1\cos^2\theta_0+\kappa_2\sin^2\theta_0.
\end{equation}
As $\cos 2\theta_0=\xi$, we have $\cos^2\theta_0=\frac12(1+\xi)$
and $\sin^2\theta_0=\frac12(1-\xi)$, so
\begin{equation}\label{eq:curvature_n}
\kappa_n=\frac12[\kappa_1+\kappa_2+\xi(\kappa_1-\kappa_2)]. 
\end{equation}
In the limiting cases of existence for this solution, it lines up with the
principal directions of curvature: when $\xi=1$, then $\kappa_n=\kappa_1$, and when $\xi=-1$, then
$\kappa_n=\kappa_2$. In the extreme case when $k_2=0$,
we have $\xi=\frac{\kappa_2+\kappa_1}{\kappa_2-\kappa_1}$, and it then follows that $\kappa_n=0$,
independent of the actual manifold: at hyperbolic points the director prefers to align along directions
of zero curvature.

In general, using \eqref{eq:xi} in \eqref{eq:curvature_n} shows that
\begin{equation}\label{eq:curvature_n_2}
\kappa_n=H\frac{k_2}{k_2-k_3}, 
\end{equation}
where $H=\frac12(\kappa_1+\kappa_2)$ is the mean curvature.
So $\kappa_n$ will normally neither be zero nor one of the principal
curvatures.

\begin{figure}[h]
\includegraphics[width=0.48\textwidth]{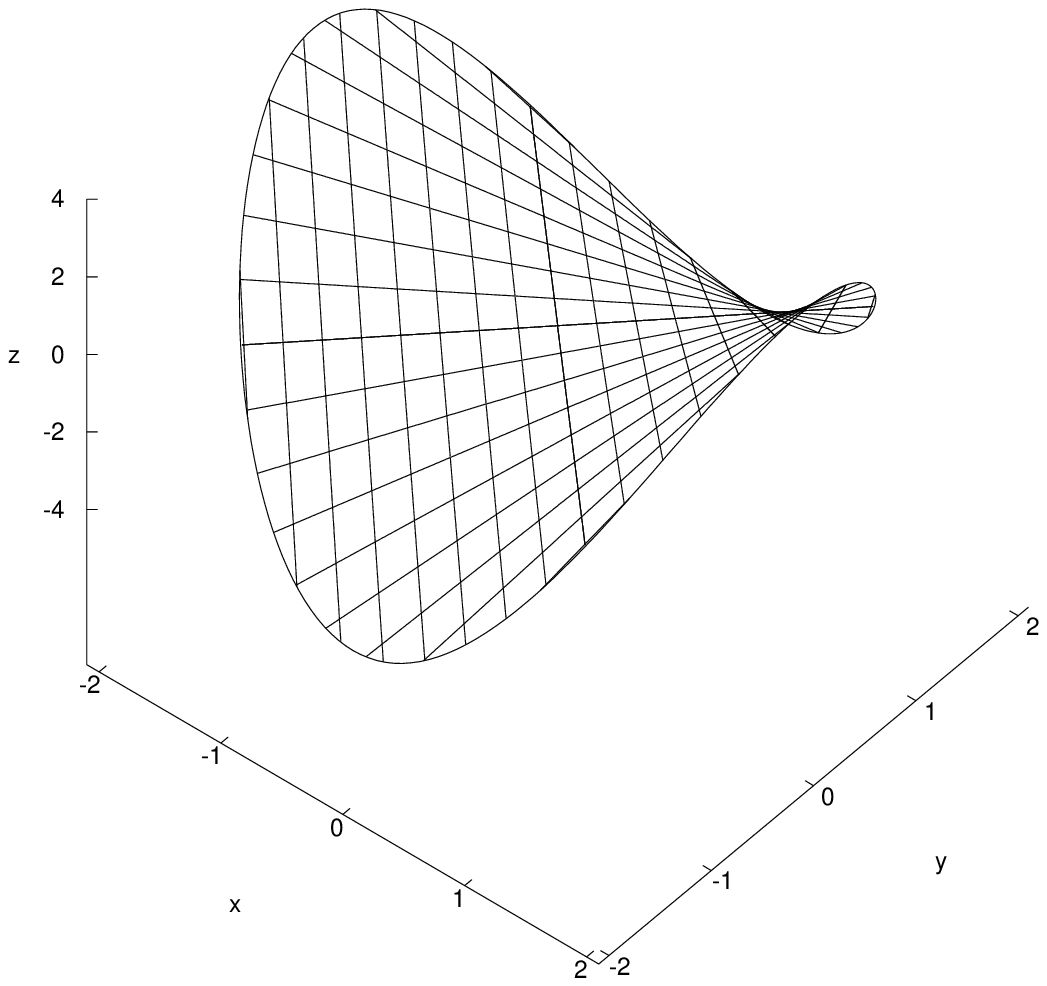}\\[10pt]
\includegraphics[width=0.48\textwidth]{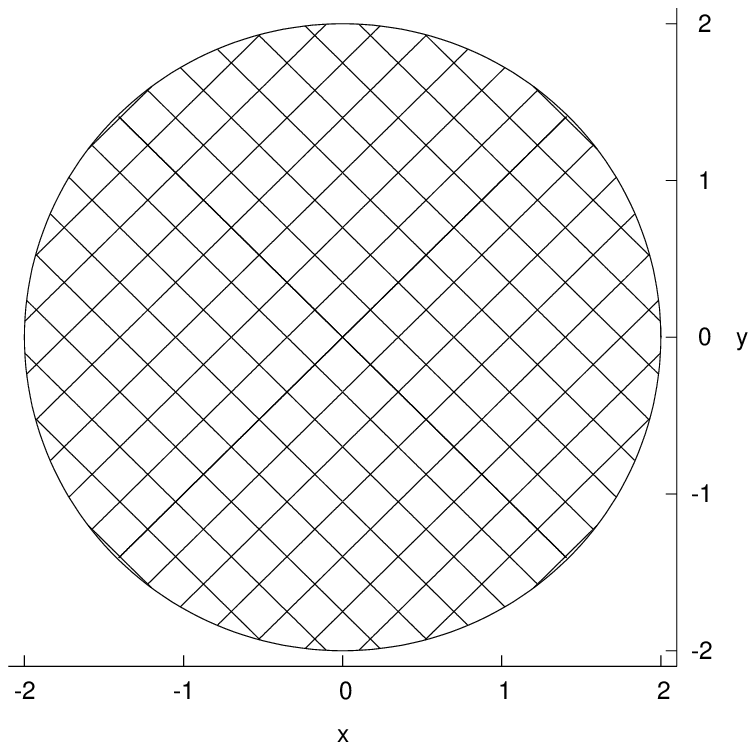}
\caption{When $k_2=0$, the director would like to align along directions of zero curvature.
In the case of the hyperbolic paraboloid, the projections of these lines onto the 
$x$-$y$-plane are lines with slope $\pm 1$.\label{fig:newLines}}
\end{figure}

\subsection{Example: hyperbolic paraboloid}\label{sec:example}
To illustrate our results, we consider as an example 
the hyperbolic paraboloid, a two-dimensional manifold which has negative
Gaussian curvature everywhere. It can be represented in the form
\begin{equation}
f(x,y)=(x,y,h(x,y))
\end{equation}
with the height of the manifold above the $x$-$y$-plane given by
$h(x,y)=x^2-y^2$. All geometric properties of the hyperbolic paraboloid, such
as the principal curvatures and principal lines of curvature, can be expressed
in terms of the height function $h$ and its derivatives~\cite[p. 137]{spivak:diffferential3}. 

The principal lines of curvature are shown in
Fig. \ref{fig:principal}, with the solid lines corresponding to the
lower fossil energy density.

In terms of polar coordinates $(r,\phi)$, the Gaussian curvature $K$ and the mean curvature $H$ are
\begin{equation}\label{eq:H_and_K}
K=-\frac{4}{(1+4r^2)^2}\quad\text{and}\quad 
H=-\frac{4r^2\cos2\phi}{(1+4r^2)^{3/2}}.
\end{equation}
As $K<0$, all points are hyperbolic, so depending on the ratio between $k_2$ and 
$k_3$ the orientation corresponding to \eqref{eq:cos2theta0} can be the one favoured
by the curvature potential. We consider the two cases $k_2=0$ and $k_2=\frac12 k_3$.

When $k_2=0$, we have $\xi=\frac{\kappa_2+\kappa_1}{\kappa_2-\kappa_1}$, and 
from \eqref{eq:xi_simple} it is evident that $0<\xi<1$ everywhere. Therefore,
the solution corresponding to \eqref{eq:cos2theta0} minimises everywhere the curvature potential
and by \eqref{eq:curvature_n_2} the director thus would align along lines with vanishing
curvature. It is known that the hyperbolic paraboloid is a
doubly-ruled surface~\cite[p. 155]{spivak:diffferential3}, 
and those rules are precisely the lines along which the curvature potential would
like to align the director, see Fig.~\ref{fig:newLines}.
Note that at every point of the manifold there are two preferred directions with
equal fossil energy density; the bistable region $\surface_0$ is the whole of $\surface$.

\begin{figure}
\includegraphics[width=0.48\textwidth]{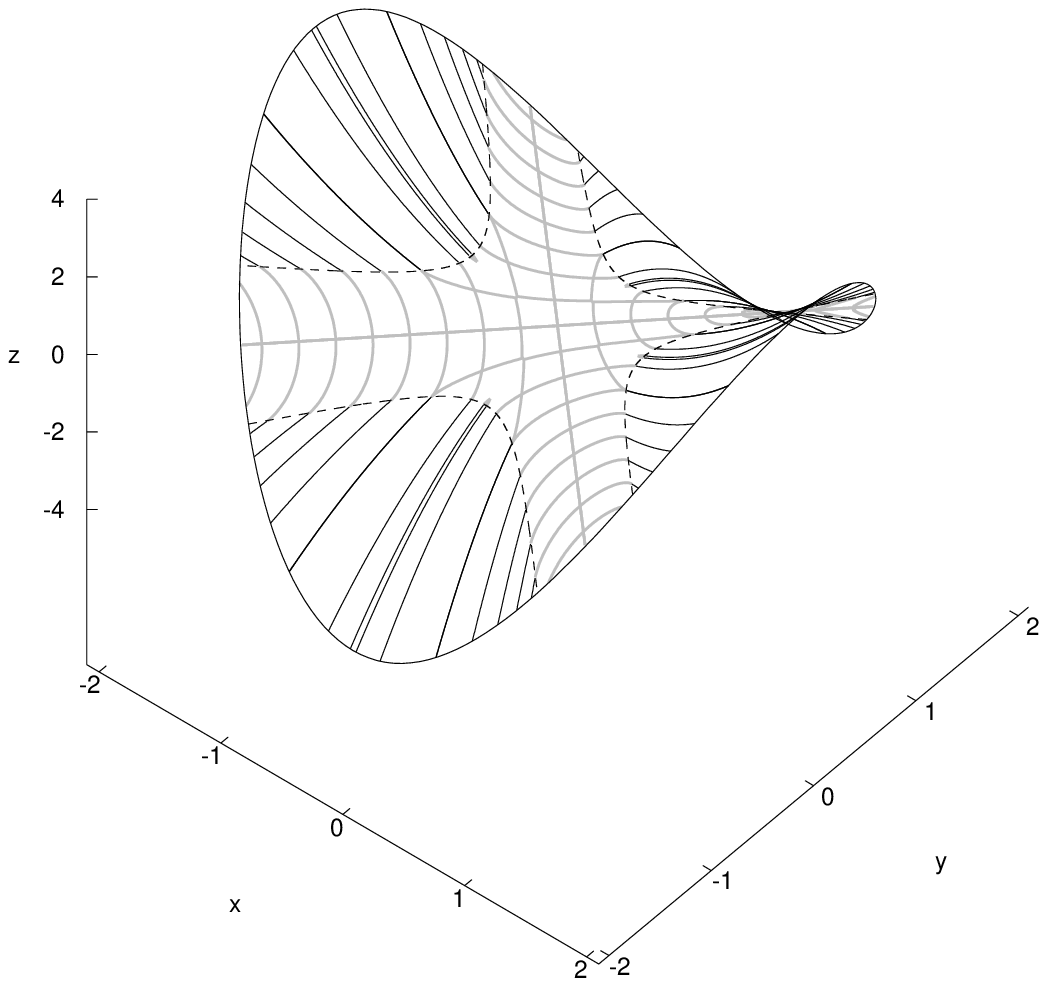}\\[10pt]
\includegraphics[width=0.48\textwidth]{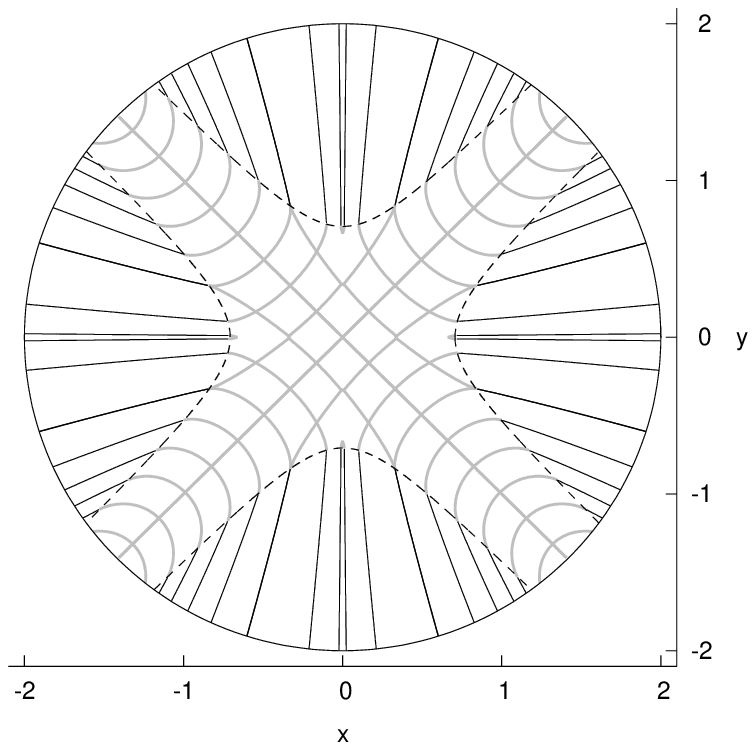}
\caption{Directions with minimal fossil energy density for $k_2=\frac12 k_3$.
Near the origin and the quadrant bisectors $0\le\xi<1$.
This is the bistable region for the fossil energy, there the two
preferred directions are depicted in grey. Where $\xi>1$, the director has only one
preferred orientation, shown in black: along the principal direction with
curvature with smaller absolute value. 
The dashed lines, along which $\xi=1$, delimit the bistable region,
according to \eqref{eq:bistable_region_delimiter_example}.
\label{fig:preferred}}
\end{figure}

When $k_2\ne0$, the orientation corresponding to \eqref{eq:cos2theta0}
only exists near the origin and near the quadrant-bisectors where $0\le\xi <1$.
In this region, there are two preferred directions, depicted in grey
in Fig.~\ref{fig:preferred}
for the case where $k_2=\frac12 k_3$,
and so $\xi=2\frac{\kappa_2+\kappa_1}{\kappa_2-\kappa_1}$. In the
same figure, the single preferred orientation in regions where $\xi>1$
is shown in black: it is along the principal direction with
curvature with smaller absolute value. In the same parametrization as in \eqref{eq:H_and_K},
\begin{equation}
\xi=\left|\frac{4r^2\cos2\phi}{\sqrt{1+4r^2+4r^4\cos^22\phi}}\right|,
\end{equation}
so that the bistable region $\surface_0$ is now characterized by
\begin{equation}\label{eq:bistable_region_delimiter_example}
r^2<\frac{2+\sqrt{4+12\cos^22\phi}}{12\cos^22\phi}.
\end{equation}

\section{Conclusions and Discussion}\label{sec:conclusions}
This paper is mainly concerned with the notion of a fossil energy that arises in nematic shells upon properly defining
the local ground state of their distortion by use of Levi-Civita's parallel transport on surfaces. From the perspective
afforded  by this notion, it is easily possible to distinguish and categorize the different contributions to the
elastic energy of nematic shells. 

Some studies, pioneered by \cite{nelson:fluctuations}  and \cite{lubensky:orientational},\footnote{Occasionally,
\cite{straley:liquid} is also  indicated as a remote, inspiring antecedent for these studies.} simply  proposed
to disregard the fossil energy and to penalize only the distortions built upon the ground state, which are
measured by the covariant gradient $\cov\n$ of the nematic director field $\n$. These are the \emph{intrinsic}
theories of nematic shells. Other studies, of which presumably the oldest is \cite{helfrich:intrinsic},
followed by \cite{kamien:extrinsic}, \cite{mbanga:frustrated}, and 
\cite{napoli2012extrinsic},%
\footnote{Contributions such as \cite{biscari:nematic,santangelo:geometric,jiang:vesicle,frank:defects} should also be mentioned,
for completeness, even if some are originating from the world of flexible vesicles, rather than from that of rigid nematic shells.}
have duly incorporated the fossil energy into the energetics of nematic shells, mostly advocating, however, a different
interpretation of its effects in terms of certain \emph{extrinsic} geometric features revealed by the flux lines of $\n$.

From the latest studies, especially on applications to generalized cylinders \cite{napoli2013curvature},
a widespread opinion has emerged, elevated to the dignity of \emph{heuristic principle} in \cite{segatti2016analysis},
which in its proponents' own words reads: ``the nematic elastic energy promotes the alignment of the flux lines
of the nematic director towards geodesics and/or lines of curvature of the surface'' \cite{napoli2012extrinsic}.
The clear-cut distinction between distortional and fossil energies, which we have based on an elementary argument,
has unambiguously identified the antagonistic players striving for a minimum. 
Clarity breeds novelty: the
fossil energy acts as a curvature potential that in general does \emph{not} orient $\n$ along the
direction with minimum absolute principal curvature, as is the case for cylinders. 
Actually, we prove in Apppendix \ref{sec:app_developable} that cylinders are  very special, as they are the only surfaces on which
the fossil energy can be uniformly minimized. \emph{Two} equivalent, but distinct fossil preferred orientations, \emph{none} of which along a principal direction
of curvature, may arise at \emph{hyperbolic points} of the shell, where the Gaussian curvature is negative.
A noticeable example is offered by a quadratic saddle surface, whose points are all hyperbolic.

It would be interesting to see whether inclusion of the fossil energy in the shape functional studied by
Giomi~\cite{giomi:hyperbolic} would affect the stability analysis (and the related shape transformation)
for the soft hyperbolic interfaces he considered. We know already from \cite{mbanga:frustrated} that
catenoids endowed with a phenomenological  \emph{ad hoc} extrinsic energy, not inherited from the fossil energy,
induce on the nematic director a generic preference to be oriented in the direction of local maximum or minimum curvature,
a preference more pronounced in regions with higher curvature anisotropy. We should ask whether
the preference for a skew fossil orientation would result in a different energy landscape.  

These are problems yet to be solved. One already tackled is the equilibrium problem for the elastic energy density
in \eqref{eq:W_general_representation_formula} on a circular toroidal shell
\cite{li:defect-free,segatti:equilibrium,segatti2016analysis} (see also \cite{bowick:curvature-induced,evans:phase}
for parallel treatments with only intrinsic, distortional energies, and \cite{ye:nematic} for an
interesting application to toroidal shells of a promising Onsager excluded volume theory for wormlike polymer
chains elaborated by Chen~\cite{chen:theory}). But, unfortunately, this problem has only been solved for 
the singular case where $k_2=k_3$,
which by \eqref{eq:fossil_energy_revealed} implies that the fossil preferred direction coincides with
the direction of minimal absolute curvature (that is, with either parallles or meridians).
Since all around the hole of the torus the Gaussian curvature is negative, for $k_2<k_3$ the fossil preferred directions
are different from both parallels and meridians. The equilibrium bifurcation pattern should arguably be richer
than anticipated for the case $k_2=k_3$, where for sufficiently fat tori the equilibrium director was found
to bend towards the meridians while swinging through the hole.

As our explicit example indicated, there may be two equivalent, but distinct  fossil preferred directions on the shell.
This should not come as a surprise, since in general the fossil energy cannot be uniformly minimized by the same orientation
in the principal curvature frame, save on generalized cylinders. One might think that bistability of the fossil energy
should herald  defect formation, in analogy to the criterion established by some intrinsic
theories according to which defects (of whatever topological charge) are attracted to regions of negative Gaussian
curvature (and repelled from regions of positive Gaussian curvature); see, in particular,
\cite{vitelli:anomalous,vitelli:defect} for learning more about this specific claim, and \cite{bowick:two-dimensional} 
for a broader, enlightening review. Though the connection between fossil energy and defect formation is worth exploring,
it may turn out to be more complicated than expected. In general, distortional  and fossil energies compete one against
the other, just because a single fossil preferred direction cannot be uniformly enforced; defects will arise wherever such a
competition cannot be smoothly resolved and the system would rather prefer paying the extra cost that in a
doctored\footnote{The divergence of the elastic energy at a point defect in a two-dimensional setting is usually tamed
by carving out a small core about the defect and attributing to it a standard melting cost \cite{lubensky:orientational}.}
two-dimensional director theory is due by a defect for the total energy to remain finite.  
\begin{acknowledgments}
	E.G.V. acknowledges the kind hospitality of the Oxford Centre for
	Nonlinear PDE, where part of this work was done while he was visiting the
	Mathematical Institute at the University of Oxford.
\end{acknowledgments}
\appendix
\section{Surfaces with minimum fossil energy}\label{sec:app_developable}
In this appendix, we address two related issues, which are perhaps of a more technical character. 

First, returning to the expression in \eqref{eq:minimum_distortion_solution} for the measure of least distortion, we ask whether $\mindis$ could indeed be the gradient of a director field, at least locally. The answer follows easily from the very parallel transport construction: for $\mindis$ to be a gradient, it must be possible to parallel transport a unit vector $\nzero$ around any closed loop $\curve$ on $\surface$ generating \emph{no} mismatch $\Delta\nzero$, that is, 
\begin{equation}\label{eq:mismatch}
\Delta\nzero=\int_{\curve}[\mindis]\tangent\dd s=\bzero,
\end{equation} 
where $\tangent$ denotes the unit tangent along $\curve$ and $s$ is the arc-length measured along it. There is an equivalent way to express this condition. If the field $\nzero$ exists, then (see equation (88) of \cite{rosso2012parallel})
\begin{equation}\label{eq:mismatch_equivalent}
\Delta\nzero=\int_{\surfaceC}K\nzero_\perp\dd a=\bzero,
\end{equation}
where $a$ denotes the area measure, $K=\kappa_1\kappa_2$ is the Gaussian curvature of $\surface$, $\nzero_\perp:=\N\nzero$, and $\surfaceC$ is the portion of $\surface$ delimited by $\curve$. Requiring \eqref{eq:mismatch_equivalent} to be satisfied for all curves amounts to require that $K\equiv0$, which means that  $\surface$ is a \emph{developable} surface. It is known (see, for example, p.~448 of \cite{gray:differential}) that a developable surface is locally either a generalized cylinder, or a generalized cone, or a tangent developable. In all these cases, we learn from \eqref{eq:grad_n_alpha_rewritten} that a minimal distortion field $\nzero$ can be globally imprinted on $\surface$ with $\alpha\equiv\mathrm{const}$ (and so $\cov\nzero\equiv\bzero$), which in particular implies that the whole distortion energy is fossil.

The second question we ask is whether in the cases where the whole distortion energy is fossil there is a class of surfaces
$\surface$ on which this energy can be minimized everywhere without introducing any extra local distortion.
To answer this question we first note that $\kappa_1=0$, since $K=0$, and from the discussion in Sec.~\ref{sec:multiplicity_bifurcation}
above it follows that $W_0$ is locally minimized by $\theta=0$. This results in a global minimum only if
the corresponding principal direction of curvature, $\e_1$ (and consequently also $\e_2$) is parallel
transported all along $\surface$. Our claim is that among all developable surfaces only generalized
cylinders enjoy this property (thus justifying the role played by these special shells in \cite{napoli2013curvature}).
We shall only prove directly that tangent developable surfaces do \emph{not} enjoy this property; a slight modification 
of the same argument would then allow the reader to conclude the proof for generalized cylinders and
generalized cones.\footnote{For these cases, a synthetic proof could also be given based on the geometric consideration that,
once straightened out in the planar developed surface, the lines of curvature of a generalized cylinder are parallel,
whereas those of a generalized cone are convergent.}

A tangent developable surface $\surface$ is represented as follows (see, for example, p.~438 of \cite{gray:differential}),
\begin{equation}\label{eq:tangent_developable_representation}
p(s,u)=\rv(s)+u\tangent(s),
\end{equation}
where $\rv(s)$ is the \emph{striction curve} of $\surface$, parametrized by arc-length $s$, $\tangent(s)$ is the unit tangent to $\rv(s)$, and $u$ is an additional parameter representing a length. Letting both $s$ and $u$ depend on a further parameter $t$, so as to describe a curve $\curve$ on $\surface$, we easily see from that \eqref{eq:tangent_developable_representation} that 
\begin{equation}\label{eq:p_dot}
\dot{p}=(\dot{s}+\dot{u})\tangent +\dot{s}u\sigma\n,
\end{equation} 
where a superimposed dot denotes differentiation with respect to $t$, $\sigma$ is the principal curvature of $\rv(s)$ and $\n$ is its principal unit normal. Exploiting in \eqref{eq:p_dot} the arbitrariness of $\curve$, we easily see that the unit normal $\normal$ to $\surface$ coincides with the binormal unit vector $\binormal$ to $\rv$, from which it follows that 
\begin{equation}\label{eq:normal_dot}
\dot{\normal}=(\ctens)\dot{p}=\tau\dot{s}\n,
\end{equation}
where $\tau$ is the torsion of $\rv(s)$. Combining \eqref{eq:normal_dot} and \eqref{eq:p_dot}, we can write
\begin{equation}\label{eq:curvature_tangent_developable}
\ctens=\frac{1}{u}\frac{\tau}{\sigma}\n\otimes\n,
\end{equation}
which shows that one principal direction of curvature is $\tangent$, with $\kappa_1=0$, as expected, and the other is $\n$, with
$\kappa_2=u/\tau\sigma$. 

To see whether $\tangent$ and $\n$ are parallel transported on $\surface$, it suffices to compute $\sgradn$. Denoting by a prime differentiation with respect to $s$, we first compute
\begin{equation}
\dot\n=(\sgradn)\dot{p}=\dot{s}\n'=-\dot{s}(\sigma\tangent+\tau\normal),
\end{equation}
from which we obtain that 
\begin{equation}\label{eq:sgradn_tangent_developable}
\sgradn=-\frac1u\n_\perp\otimes\n-\frac1u\frac{\tau}{\sigma}\normal\otimes\n.
\end{equation}
By \eqref{eq:curvature_tangent_developable} and \eqref{eq:minimum_distortion_solution}, we can also write the last term in \eqref{eq:sgradn_tangent_developable} as
\begin{equation}
-\frac1u\frac{\tau}{\sigma}\normal\otimes\n=\mindis,
\end{equation}
thus concluding that $\n$ is \emph{not} parallel transported on $\surface$, since $\sgradn\neq\mindis$.



\bibliography{curvature_EGV.bib}

\begin{thebibliography}{67}%
\makeatletter
\providecommand \@ifxundefined [1]{%
 \@ifx{#1\undefined}
}%
\providecommand \@ifnum [1]{%
 \ifnum #1\expandafter \@firstoftwo
 \else \expandafter \@secondoftwo
 \fi
}%
\providecommand \@ifx [1]{%
 \ifx #1\expandafter \@firstoftwo
 \else \expandafter \@secondoftwo
 \fi
}%
\providecommand \natexlab [1]{#1}%
\providecommand \enquote  [1]{``#1''}%
\providecommand \bibnamefont  [1]{#1}%
\providecommand \bibfnamefont [1]{#1}%
\providecommand \citenamefont [1]{#1}%
\providecommand \href@noop [0]{\@secondoftwo}%
\providecommand \href [0]{\begingroup \@sanitize@url \@href}%
\providecommand \@href[1]{\@@startlink{#1}\@@href}%
\providecommand \@@href[1]{\endgroup#1\@@endlink}%
\providecommand \@sanitize@url [0]{\catcode `\\12\catcode `\$12\catcode
  `\&12\catcode `\#12\catcode `\^12\catcode `\_12\catcode `\%12\relax}%
\providecommand \@@startlink[1]{}%
\providecommand \@@endlink[0]{}%
\providecommand \url  [0]{\begingroup\@sanitize@url \@url }%
\providecommand \@url [1]{\endgroup\@href {#1}{\urlprefix }}%
\providecommand \urlprefix  [0]{URL }%
\providecommand \Eprint [0]{\href }%
\providecommand \doibase [0]{http://dx.doi.org/}%
\providecommand \selectlanguage [0]{\@gobble}%
\providecommand \bibinfo  [0]{\@secondoftwo}%
\providecommand \bibfield  [0]{\@secondoftwo}%
\providecommand \translation [1]{[#1]}%
\providecommand \BibitemOpen [0]{}%
\providecommand \bibitemStop [0]{}%
\providecommand \bibitemNoStop [0]{.\EOS\space}%
\providecommand \EOS [0]{\spacefactor3000\relax}%
\providecommand \BibitemShut  [1]{\csname bibitem#1\endcsname}%
\let\auto@bib@innerbib\@empty
\bibitem [{\citenamefont {Arsenault}\ \emph {et~al.}(2004)\citenamefont
  {Arsenault}, \citenamefont {Fournier-Bidoz}, \citenamefont {Hatton},
  \citenamefont {Miguez}, \citenamefont {Tetreault}, \citenamefont {Vekris},
  \citenamefont {Wong}, \citenamefont {Ming~Yang}, \citenamefont {Kitaev},\
  and\ \citenamefont {Ozin}}]{arsenault:towards}%
  \BibitemOpen
  \bibfield  {author} {\bibinfo {author} {\bibfnamefont {A.}~\bibnamefont
  {Arsenault}}, \bibinfo {author} {\bibfnamefont {S.}~\bibnamefont
  {Fournier-Bidoz}}, \bibinfo {author} {\bibfnamefont {B.}~\bibnamefont
  {Hatton}}, \bibinfo {author} {\bibfnamefont {H.}~\bibnamefont {Miguez}},
  \bibinfo {author} {\bibfnamefont {N.}~\bibnamefont {Tetreault}}, \bibinfo
  {author} {\bibfnamefont {E.}~\bibnamefont {Vekris}}, \bibinfo {author}
  {\bibfnamefont {S.}~\bibnamefont {Wong}}, \bibinfo {author} {\bibfnamefont
  {S.}~\bibnamefont {Ming~Yang}}, \bibinfo {author} {\bibfnamefont
  {V.}~\bibnamefont {Kitaev}}, \ and\ \bibinfo {author} {\bibfnamefont {G.~A.}\
  \bibnamefont {Ozin}},\ }\href@noop {} {\bibfield  {journal} {\bibinfo
  {journal} {J. Mater. Chem.}\ }\textbf {\bibinfo {volume} {14}},\ \bibinfo
  {pages} {781} (\bibinfo {year} {2004})}\BibitemShut {NoStop}%
\bibitem [{\citenamefont {Nelson}(2002)}]{nelson:toward}%
  \BibitemOpen
  \bibfield  {author} {\bibinfo {author} {\bibfnamefont {D.~R.}\ \bibnamefont
  {Nelson}},\ }\href@noop {} {\bibfield  {journal} {\bibinfo  {journal} {Nano
  Lett.}\ }\textbf {\bibinfo {volume} {2}},\ \bibinfo {pages} {1125} (\bibinfo
  {year} {2002})}\BibitemShut {NoStop}%
\bibitem [{\citenamefont {Huber}\ and\ \citenamefont
  {Stark}(2005)}]{huber:tetravalent}%
  \BibitemOpen
  \bibfield  {author} {\bibinfo {author} {\bibfnamefont {M.}~\bibnamefont
  {Huber}}\ and\ \bibinfo {author} {\bibfnamefont {H.}~\bibnamefont {Stark}},\
  }\href@noop {} {\bibfield  {journal} {\bibinfo  {journal} {Europhys. Lett.}\
  }\textbf {\bibinfo {volume} {69}},\ \bibinfo {pages} {135} (\bibinfo {year}
  {2005})}\BibitemShut {NoStop}%
\bibitem [{\citenamefont {Fern\'andez-Nieves}\ \emph
  {et~al.}(2007)\citenamefont {Fern\'andez-Nieves}, \citenamefont {Vitelli},
  \citenamefont {Utada}, \citenamefont {Link}, \citenamefont {M\'arquez},
  \citenamefont {Nelson},\ and\ \citenamefont
  {Weitz}}]{fernandez-nieves:novel}%
  \BibitemOpen
  \bibfield  {author} {\bibinfo {author} {\bibfnamefont {A.}~\bibnamefont
  {Fern\'andez-Nieves}}, \bibinfo {author} {\bibfnamefont {V.}~\bibnamefont
  {Vitelli}}, \bibinfo {author} {\bibfnamefont {A.~S.}\ \bibnamefont {Utada}},
  \bibinfo {author} {\bibfnamefont {D.~R.}\ \bibnamefont {Link}}, \bibinfo
  {author} {\bibfnamefont {M.}~\bibnamefont {M\'arquez}}, \bibinfo {author}
  {\bibfnamefont {D.~R.}\ \bibnamefont {Nelson}}, \ and\ \bibinfo {author}
  {\bibfnamefont {D.~A.}\ \bibnamefont {Weitz}},\ }\href@noop {} {\bibfield
  {journal} {\bibinfo  {journal} {Phys. Rev. Lett.}\ }\textbf {\bibinfo
  {volume} {99}},\ \bibinfo {pages} {157801} (\bibinfo {year}
  {2007})}\BibitemShut {NoStop}%
\bibitem [{\citenamefont {Lopez-Leon}\ \emph
  {et~al.}(2011{\natexlab{a}})\citenamefont {Lopez-Leon}, \citenamefont
  {Koning}, \citenamefont {Devaiah}, \citenamefont {Vitelli},\ and\
  \citenamefont {Fernandez-Nieves}}]{lopez-leon:frustrated}%
  \BibitemOpen
  \bibfield  {author} {\bibinfo {author} {\bibfnamefont {T.}~\bibnamefont
  {Lopez-Leon}}, \bibinfo {author} {\bibfnamefont {V.}~\bibnamefont {Koning}},
  \bibinfo {author} {\bibfnamefont {K.~B.~S.}\ \bibnamefont {Devaiah}},
  \bibinfo {author} {\bibfnamefont {V.}~\bibnamefont {Vitelli}}, \ and\
  \bibinfo {author} {\bibfnamefont {A.}~\bibnamefont {Fernandez-Nieves}},\
  }\href@noop {} {\bibfield  {journal} {\bibinfo  {journal} {Nature Phys.}\
  }\textbf {\bibinfo {volume} {7}},\ \bibinfo {pages} {391} (\bibinfo {year}
  {2011}{\natexlab{a}})}\BibitemShut {NoStop}%
\bibitem [{\citenamefont {Lopez-Leon}\ \emph
  {et~al.}(2011{\natexlab{b}})\citenamefont {Lopez-Leon}, \citenamefont
  {Fernandez-Nieves}, \citenamefont {Nobili},\ and\ \citenamefont
  {Blanc}}]{lopez-leon:nematic-smectic}%
  \BibitemOpen
  \bibfield  {author} {\bibinfo {author} {\bibfnamefont {T.}~\bibnamefont
  {Lopez-Leon}}, \bibinfo {author} {\bibfnamefont {A.}~\bibnamefont
  {Fernandez-Nieves}}, \bibinfo {author} {\bibfnamefont {M.}~\bibnamefont
  {Nobili}}, \ and\ \bibinfo {author} {\bibfnamefont {C.}~\bibnamefont
  {Blanc}},\ }\href@noop {} {\bibfield  {journal} {\bibinfo  {journal} {Phys.
  Rev. Lett.}\ }\textbf {\bibinfo {volume} {106}},\ \bibinfo {pages} {247802}
  (\bibinfo {year} {2011}{\natexlab{b}})}\BibitemShut {NoStop}%
\bibitem [{\citenamefont {Liang}\ \emph {et~al.}(2011)\citenamefont {Liang},
  \citenamefont {Schymura}, \citenamefont {Rudquist},\ and\ \citenamefont
  {Lagerwall}}]{liang:nematic-smectic}%
  \BibitemOpen
  \bibfield  {author} {\bibinfo {author} {\bibfnamefont {H.-L.}\ \bibnamefont
  {Liang}}, \bibinfo {author} {\bibfnamefont {S.}~\bibnamefont {Schymura}},
  \bibinfo {author} {\bibfnamefont {P.}~\bibnamefont {Rudquist}}, \ and\
  \bibinfo {author} {\bibfnamefont {J.}~\bibnamefont {Lagerwall}},\ }\href@noop
  {} {\bibfield  {journal} {\bibinfo  {journal} {Phys. Rev. Lett.}\ }\textbf
  {\bibinfo {volume} {106}},\ \bibinfo {pages} {247801} (\bibinfo {year}
  {2011})}\BibitemShut {NoStop}%
\bibitem [{\citenamefont {Se\v{c}}\ \emph {et~al.}(2012)\citenamefont
  {Se\v{c}}, \citenamefont {Lopez-Leon}, \citenamefont {Nobili}, \citenamefont
  {Blanc}, \citenamefont {Fernandez-Nieves}, \citenamefont {Ravnik},\ and\
  \citenamefont {\v{Z}umer}}]{sec:defect}%
  \BibitemOpen
  \bibfield  {author} {\bibinfo {author} {\bibfnamefont {D.}~\bibnamefont
  {Se\v{c}}}, \bibinfo {author} {\bibfnamefont {T.}~\bibnamefont {Lopez-Leon}},
  \bibinfo {author} {\bibfnamefont {M.}~\bibnamefont {Nobili}}, \bibinfo
  {author} {\bibfnamefont {C.}~\bibnamefont {Blanc}}, \bibinfo {author}
  {\bibfnamefont {A.}~\bibnamefont {Fernandez-Nieves}}, \bibinfo {author}
  {\bibfnamefont {M.}~\bibnamefont {Ravnik}}, \ and\ \bibinfo {author}
  {\bibfnamefont {S.}~\bibnamefont {\v{Z}umer}},\ }\href@noop {} {\bibfield
  {journal} {\bibinfo  {journal} {Phys. Rev. E}\ }\textbf {\bibinfo {volume}
  {86}},\ \bibinfo {pages} {020705} (\bibinfo {year} {2012})}\BibitemShut
  {NoStop}%
\bibitem [{\citenamefont {Lopez-Leon}\ \emph {et~al.}(2012)\citenamefont
  {Lopez-Leon}, \citenamefont {Bates},\ and\ \citenamefont
  {Fernandez-Nieves}}]{lopez-leon:defect}%
  \BibitemOpen
  \bibfield  {author} {\bibinfo {author} {\bibfnamefont {T.}~\bibnamefont
  {Lopez-Leon}}, \bibinfo {author} {\bibfnamefont {M.~A.}\ \bibnamefont
  {Bates}}, \ and\ \bibinfo {author} {\bibfnamefont {A.}~\bibnamefont
  {Fernandez-Nieves}},\ }\href@noop {} {\bibfield  {journal} {\bibinfo
  {journal} {Phys. Rev. E}\ }\textbf {\bibinfo {volume} {86}},\ \bibinfo
  {pages} {030702} (\bibinfo {year} {2012})}\BibitemShut {NoStop}%
\bibitem [{\citenamefont {Noh}\ \emph {et~al.}(2016{\natexlab{a}})\citenamefont
  {Noh}, \citenamefont {Henx},\ and\ \citenamefont {Lagerwall}}]{noh:taming}%
  \BibitemOpen
  \bibfield  {author} {\bibinfo {author} {\bibfnamefont {J.}~\bibnamefont
  {Noh}}, \bibinfo {author} {\bibfnamefont {B.}~\bibnamefont {Henx}}, \ and\
  \bibinfo {author} {\bibfnamefont {J.~P.~F.}\ \bibnamefont {Lagerwall}},\
  }\href@noop {} {\bibfield  {journal} {\bibinfo  {journal} {Adv. Mater.}\
  }\textbf {\bibinfo {volume} {28}},\ \bibinfo {pages} {10170} (\bibinfo {year}
  {2016}{\natexlab{a}})}\BibitemShut {NoStop}%
\bibitem [{\citenamefont {Noh}\ \emph {et~al.}(2016{\natexlab{b}})\citenamefont
  {Noh}, \citenamefont {Reguengo De~Sousa},\ and\ \citenamefont
  {Lagerwall}}]{noh:influence}%
  \BibitemOpen
  \bibfield  {author} {\bibinfo {author} {\bibfnamefont {J.}~\bibnamefont
  {Noh}}, \bibinfo {author} {\bibfnamefont {K.}~\bibnamefont {Reguengo
  De~Sousa}}, \ and\ \bibinfo {author} {\bibfnamefont {J.~P.~F.}\ \bibnamefont
  {Lagerwall}},\ }\href@noop {} {\bibfield  {journal} {\bibinfo  {journal}
  {Soft Matter}\ }\textbf {\bibinfo {volume} {12}},\ \bibinfo {pages} {367}
  (\bibinfo {year} {2016}{\natexlab{b}})}\BibitemShut {NoStop}%
\bibitem [{\citenamefont {Koning}\ \emph {et~al.}(2016)\citenamefont {Koning},
  \citenamefont {Lopez-Leon}, \citenamefont {Darmon}, \citenamefont
  {Fernandez-Nieves},\ and\ \citenamefont {Vitelli}}]{koning:spherical}%
  \BibitemOpen
  \bibfield  {author} {\bibinfo {author} {\bibfnamefont {V.}~\bibnamefont
  {Koning}}, \bibinfo {author} {\bibfnamefont {T.}~\bibnamefont {Lopez-Leon}},
  \bibinfo {author} {\bibfnamefont {A.}~\bibnamefont {Darmon}}, \bibinfo
  {author} {\bibfnamefont {A.}~\bibnamefont {Fernandez-Nieves}}, \ and\
  \bibinfo {author} {\bibfnamefont {V.}~\bibnamefont {Vitelli}},\ }\href@noop
  {} {\bibfield  {journal} {\bibinfo  {journal} {Phys. Rev. E}\ }\textbf
  {\bibinfo {volume} {94}},\ \bibinfo {pages} {012703} (\bibinfo {year}
  {2016})}\BibitemShut {NoStop}%
\bibitem [{\citenamefont {Ska\v{c}ej}\ and\ \citenamefont
  {Zannoni}(2008)}]{skacej:controlling}%
  \BibitemOpen
  \bibfield  {author} {\bibinfo {author} {\bibfnamefont {G.}~\bibnamefont
  {Ska\v{c}ej}}\ and\ \bibinfo {author} {\bibfnamefont {C.}~\bibnamefont
  {Zannoni}},\ }\href@noop {} {\bibfield  {journal} {\bibinfo  {journal} {Phys.
  Rev. Lett.}\ }\textbf {\bibinfo {volume} {100}},\ \bibinfo {pages} {197802}
  (\bibinfo {year} {2008})}\BibitemShut {NoStop}%
\bibitem [{\citenamefont {Shin}\ \emph {et~al.}(2008)\citenamefont {Shin},
  \citenamefont {Bowick},\ and\ \citenamefont {Xing}}]{shin:topological}%
  \BibitemOpen
  \bibfield  {author} {\bibinfo {author} {\bibfnamefont {H.}~\bibnamefont
  {Shin}}, \bibinfo {author} {\bibfnamefont {M.~J.}\ \bibnamefont {Bowick}}, \
  and\ \bibinfo {author} {\bibfnamefont {X.}~\bibnamefont {Xing}},\ }\href@noop
  {} {\bibfield  {journal} {\bibinfo  {journal} {Phys. Rev. Lett.}\ }\textbf
  {\bibinfo {volume} {101}},\ \bibinfo {pages} {037802} (\bibinfo {year}
  {2008})}\BibitemShut {NoStop}%
\bibitem [{\citenamefont {Bates}\ \emph {et~al.}(2010)\citenamefont {Bates},
  \citenamefont {Skacej},\ and\ \citenamefont {Zannoni}}]{bates:defects}%
  \BibitemOpen
  \bibfield  {author} {\bibinfo {author} {\bibfnamefont {M.~A.}\ \bibnamefont
  {Bates}}, \bibinfo {author} {\bibfnamefont {G.}~\bibnamefont {Skacej}}, \
  and\ \bibinfo {author} {\bibfnamefont {C.}~\bibnamefont {Zannoni}},\
  }\href@noop {} {\bibfield  {journal} {\bibinfo  {journal} {Soft Matter}\
  }\textbf {\bibinfo {volume} {6}},\ \bibinfo {pages} {655} (\bibinfo {year}
  {2010})}\BibitemShut {NoStop}%
\bibitem [{\citenamefont {Dhakal}\ \emph {et~al.}(2012)\citenamefont {Dhakal},
  \citenamefont {Solis},\ and\ \citenamefont {Olvera de~la
  Cruz}}]{dhakal:nematic}%
  \BibitemOpen
  \bibfield  {author} {\bibinfo {author} {\bibfnamefont {S.}~\bibnamefont
  {Dhakal}}, \bibinfo {author} {\bibfnamefont {F.~J.}\ \bibnamefont {Solis}}, \
  and\ \bibinfo {author} {\bibfnamefont {M.}~\bibnamefont {Olvera de~la
  Cruz}},\ }\href@noop {} {\bibfield  {journal} {\bibinfo  {journal} {Phys.
  Rev. E}\ }\textbf {\bibinfo {volume} {86}},\ \bibinfo {pages} {011709}
  (\bibinfo {year} {2012})}\BibitemShut {NoStop}%
\bibitem [{\citenamefont {Li}\ \emph {et~al.}(2013)\citenamefont {Li},
  \citenamefont {Miao}, \citenamefont {Ma},\ and\ \citenamefont
  {Chen}}]{li:topological}%
  \BibitemOpen
  \bibfield  {author} {\bibinfo {author} {\bibfnamefont {Y.}~\bibnamefont
  {Li}}, \bibinfo {author} {\bibfnamefont {H.}~\bibnamefont {Miao}}, \bibinfo
  {author} {\bibfnamefont {H.}~\bibnamefont {Ma}}, \ and\ \bibinfo {author}
  {\bibfnamefont {J.~Z.~Y.}\ \bibnamefont {Chen}},\ }\href@noop {} {\bibfield
  {journal} {\bibinfo  {journal} {Soft Matter}\ }\textbf {\bibinfo {volume}
  {9}},\ \bibinfo {pages} {11461} (\bibinfo {year} {2013})}\BibitemShut
  {NoStop}%
\bibitem [{\citenamefont {Li}\ \emph {et~al.}(2014)\citenamefont {Li},
  \citenamefont {Miao}, \citenamefont {Ma},\ and\ \citenamefont
  {Chen}}]{li:defect-free}%
  \BibitemOpen
  \bibfield  {author} {\bibinfo {author} {\bibfnamefont {Y.}~\bibnamefont
  {Li}}, \bibinfo {author} {\bibfnamefont {H.}~\bibnamefont {Miao}}, \bibinfo
  {author} {\bibfnamefont {H.}~\bibnamefont {Ma}}, \ and\ \bibinfo {author}
  {\bibfnamefont {J.~Z.~Y.}\ \bibnamefont {Chen}},\ }\href@noop {} {\bibfield
  {journal} {\bibinfo  {journal} {RSC Adv.}\ }\textbf {\bibinfo {volume} {4}},\
  \bibinfo {pages} {27471} (\bibinfo {year} {2014})}\BibitemShut {NoStop}%
\bibitem [{\citenamefont {Mbanga}\ \emph {et~al.}(2014)\citenamefont {Mbanga},
  \citenamefont {Voorhes},\ and\ \citenamefont {Atherton}}]{mbanga:simulating}%
  \BibitemOpen
  \bibfield  {author} {\bibinfo {author} {\bibfnamefont {B.~L.}\ \bibnamefont
  {Mbanga}}, \bibinfo {author} {\bibfnamefont {K.~K.}\ \bibnamefont {Voorhes}},
  \ and\ \bibinfo {author} {\bibfnamefont {T.~J.}\ \bibnamefont {Atherton}},\
  }\href@noop {} {\bibfield  {journal} {\bibinfo  {journal} {Phys. Rev. E}\
  }\textbf {\bibinfo {volume} {89}},\ \bibinfo {pages} {052504} (\bibinfo
  {year} {2014})}\BibitemShut {NoStop}%
\bibitem [{\citenamefont {Wand}\ and\ \citenamefont
  {Bates}(2015)}]{wand:monte}%
  \BibitemOpen
  \bibfield  {author} {\bibinfo {author} {\bibfnamefont {C.~R.}\ \bibnamefont
  {Wand}}\ and\ \bibinfo {author} {\bibfnamefont {M.~A.}\ \bibnamefont
  {Bates}},\ }\href@noop {} {\bibfield  {journal} {\bibinfo  {journal} {Phys.
  Rev. E}\ }\textbf {\bibinfo {volume} {91}},\ \bibinfo {pages} {012502}
  (\bibinfo {year} {2015})}\BibitemShut {NoStop}%
\bibitem [{\citenamefont {Kralj}\ \emph {et~al.}(2011)\citenamefont {Kralj},
  \citenamefont {Rosso},\ and\ \citenamefont {Virga}}]{kralj2011curvature}%
  \BibitemOpen
  \bibfield  {author} {\bibinfo {author} {\bibfnamefont {S.}~\bibnamefont
  {Kralj}}, \bibinfo {author} {\bibfnamefont {R.}~\bibnamefont {Rosso}}, \ and\
  \bibinfo {author} {\bibfnamefont {E.~G.}\ \bibnamefont {Virga}},\ }\href@noop
  {} {\bibfield  {journal} {\bibinfo  {journal} {Soft Matter}\ }\textbf
  {\bibinfo {volume} {7}},\ \bibinfo {pages} {670} (\bibinfo {year}
  {2011})}\BibitemShut {NoStop}%
\bibitem [{\citenamefont {Napoli}\ and\ \citenamefont
  {Vergori}(2013)}]{napoli2013curvature}%
  \BibitemOpen
  \bibfield  {author} {\bibinfo {author} {\bibfnamefont {G.}~\bibnamefont
  {Napoli}}\ and\ \bibinfo {author} {\bibfnamefont {L.}~\bibnamefont
  {Vergori}},\ }\href@noop {} {\bibfield  {journal} {\bibinfo  {journal} {Int.
  J. Non-Linear Mech.}\ }\textbf {\bibinfo {volume} {49}},\ \bibinfo {pages}
  {66} (\bibinfo {year} {2013})}\BibitemShut {NoStop}%
\bibitem [{\citenamefont {Jesenek}\ \emph {et~al.}(2015)\citenamefont
  {Jesenek}, \citenamefont {Kralj}, \citenamefont {Rosso},\ and\ \citenamefont
  {Virga}}]{jesenek:defect}%
  \BibitemOpen
  \bibfield  {author} {\bibinfo {author} {\bibfnamefont {D.}~\bibnamefont
  {Jesenek}}, \bibinfo {author} {\bibfnamefont {S.}~\bibnamefont {Kralj}},
  \bibinfo {author} {\bibfnamefont {R.}~\bibnamefont {Rosso}}, \ and\ \bibinfo
  {author} {\bibfnamefont {E.~G.}\ \bibnamefont {Virga}},\ }\href@noop {}
  {\bibfield  {journal} {\bibinfo  {journal} {Soft Matter}\ }\textbf {\bibinfo
  {volume} {11}},\ \bibinfo {pages} {2434} (\bibinfo {year}
  {2015})}\BibitemShut {NoStop}%
\bibitem [{\citenamefont {Mesarec}\ \emph {et~al.}(2016)\citenamefont
  {Mesarec}, \citenamefont {G\'o\'zd\'z}, \citenamefont {Igli\v{c}},\ and\
  \citenamefont {Kralj}}]{mesarec:effective}%
  \BibitemOpen
  \bibfield  {author} {\bibinfo {author} {\bibfnamefont {L.}~\bibnamefont
  {Mesarec}}, \bibinfo {author} {\bibfnamefont {W.}~\bibnamefont
  {G\'o\'zd\'z}}, \bibinfo {author} {\bibfnamefont {A.}~\bibnamefont
  {Igli\v{c}}}, \ and\ \bibinfo {author} {\bibfnamefont {S.}~\bibnamefont
  {Kralj}},\ }\href@noop {} {\bibfield  {journal} {\bibinfo  {journal} {Sci.
  Rep.}\ }\textbf {\bibinfo {volume} {6}},\ \bibinfo {pages} {27117} (\bibinfo
  {year} {2016})}\BibitemShut {NoStop}%
\bibitem [{\citenamefont {Zhang}\ \emph
  {et~al.}(2012{\natexlab{a}})\citenamefont {Zhang}, \citenamefont {Jiang},\
  and\ \citenamefont {Chen}}]{zhang:onsager}%
  \BibitemOpen
  \bibfield  {author} {\bibinfo {author} {\bibfnamefont {W.-Y.}\ \bibnamefont
  {Zhang}}, \bibinfo {author} {\bibfnamefont {Y.}~\bibnamefont {Jiang}}, \ and\
  \bibinfo {author} {\bibfnamefont {J.~Z.~Y.}\ \bibnamefont {Chen}},\
  }\href@noop {} {\bibfield  {journal} {\bibinfo  {journal} {Phys. Rev. Lett.}\
  }\textbf {\bibinfo {volume} {108}},\ \bibinfo {pages} {057801} (\bibinfo
  {year} {2012}{\natexlab{a}})}\BibitemShut {NoStop}%
\bibitem [{\citenamefont {Zhang}\ \emph
  {et~al.}(2012{\natexlab{b}})\citenamefont {Zhang}, \citenamefont {Jiang},\
  and\ \citenamefont {Chen}}]{zhang:solution}%
  \BibitemOpen
  \bibfield  {author} {\bibinfo {author} {\bibfnamefont {W.-Y.}\ \bibnamefont
  {Zhang}}, \bibinfo {author} {\bibfnamefont {Y.}~\bibnamefont {Jiang}}, \ and\
  \bibinfo {author} {\bibfnamefont {J.~Z.~Y.}\ \bibnamefont {Chen}},\
  }\href@noop {} {\bibfield  {journal} {\bibinfo  {journal} {Phys. Rev. E}\
  }\textbf {\bibinfo {volume} {85}},\ \bibinfo {pages} {061710} (\bibinfo
  {year} {2012}{\natexlab{b}})}\BibitemShut {NoStop}%
\bibitem [{\citenamefont {Liang}\ \emph {et~al.}(2014)\citenamefont {Liang},
  \citenamefont {Ye}, \citenamefont {Zhang},\ and\ \citenamefont
  {Chen}}]{liang:rigid}%
  \BibitemOpen
  \bibfield  {author} {\bibinfo {author} {\bibfnamefont {Q.}~\bibnamefont
  {Liang}}, \bibinfo {author} {\bibfnamefont {S.}~\bibnamefont {Ye}}, \bibinfo
  {author} {\bibfnamefont {P.}~\bibnamefont {Zhang}}, \ and\ \bibinfo {author}
  {\bibfnamefont {J.~Z.~Y.}\ \bibnamefont {Chen}},\ }\href@noop {} {\bibfield
  {journal} {\bibinfo  {journal} {J. Chem. Phys.}\ }\textbf {\bibinfo {volume}
  {141}},\ \bibinfo {pages} {244901} (\bibinfo {year} {2014})}\BibitemShut
  {NoStop}%
\bibitem [{\citenamefont {Ye}\ \emph {et~al.}(2016)\citenamefont {Ye},
  \citenamefont {Zhang},\ and\ \citenamefont {Chen}}]{ye:nematic}%
  \BibitemOpen
  \bibfield  {author} {\bibinfo {author} {\bibfnamefont {S.}~\bibnamefont
  {Ye}}, \bibinfo {author} {\bibfnamefont {P.}~\bibnamefont {Zhang}}, \ and\
  \bibinfo {author} {\bibfnamefont {J.~Z.~Y.}\ \bibnamefont {Chen}},\
  }\href@noop {} {\bibfield  {journal} {\bibinfo  {journal} {Soft Matter}\
  }\textbf {\bibinfo {volume} {12}},\ \bibinfo {pages} {5438} (\bibinfo {year}
  {2016})}\BibitemShut {NoStop}%
\bibitem [{\citenamefont {Lopez-Leon}\ and\ \citenamefont
  {Fernandez-Nieves}(2011)}]{lopez-leon:drops}%
  \BibitemOpen
  \bibfield  {author} {\bibinfo {author} {\bibfnamefont {T.}~\bibnamefont
  {Lopez-Leon}}\ and\ \bibinfo {author} {\bibfnamefont {A.}~\bibnamefont
  {Fernandez-Nieves}},\ }\href@noop {} {\bibfield  {journal} {\bibinfo
  {journal} {Colloid Polym. Sci.}\ }\textbf {\bibinfo {volume} {289}},\
  \bibinfo {pages} {345} (\bibinfo {year} {2011})}\BibitemShut {NoStop}%
\bibitem [{\citenamefont {Lagerwall}\ and\ \citenamefont
  {Scalia}(2012)}]{lagerwall:new}%
  \BibitemOpen
  \bibfield  {author} {\bibinfo {author} {\bibfnamefont {J.~P.}\ \bibnamefont
  {Lagerwall}}\ and\ \bibinfo {author} {\bibfnamefont {G.}~\bibnamefont
  {Scalia}},\ }\href@noop {} {\bibfield  {journal} {\bibinfo  {journal}
  {Current Appl. Phys.}\ }\textbf {\bibinfo {volume} {12}},\ \bibinfo {pages}
  {1387} (\bibinfo {year} {2012})}\BibitemShut {NoStop}%
\bibitem [{\citenamefont {Mirantsev}\ \emph {et~al.}(2016)\citenamefont
  {Mirantsev}, \citenamefont {de~Oliveira}, \citenamefont {de~Oliveira},\ and\
  \citenamefont {Lyra}}]{mirantsev:defect}%
  \BibitemOpen
  \bibfield  {author} {\bibinfo {author} {\bibfnamefont {L.~V.}\ \bibnamefont
  {Mirantsev}}, \bibinfo {author} {\bibfnamefont {E.~J.~L.}\ \bibnamefont
  {de~Oliveira}}, \bibinfo {author} {\bibfnamefont {I.~N.}\ \bibnamefont
  {de~Oliveira}}, \ and\ \bibinfo {author} {\bibfnamefont {M.~L.}\ \bibnamefont
  {Lyra}},\ }\href@noop {} {\bibfield  {journal} {\bibinfo  {journal} {Liquid
  Cryst. Rev.}\ }\textbf {\bibinfo {volume} {4}},\ \bibinfo {pages} {35}
  (\bibinfo {year} {2016})}\BibitemShut {NoStop}%
\bibitem [{\citenamefont {Serra}(2016)}]{serra2016curvature}%
  \BibitemOpen
  \bibfield  {author} {\bibinfo {author} {\bibfnamefont {F.}~\bibnamefont
  {Serra}},\ }\href@noop {} {\bibfield  {journal} {\bibinfo  {journal} {Liq.
  Cryst.}\ }\textbf {\bibinfo {volume} {43}},\ \bibinfo {pages} {1920}
  (\bibinfo {year} {2016})}\BibitemShut {NoStop}%
\bibitem [{\citenamefont {Urbanski}\ \emph {et~al.}(2017)\citenamefont
  {Urbanski}, \citenamefont {Reyes}, \citenamefont {Noh}, \citenamefont
  {Sharma}, \citenamefont {Geng}, \citenamefont {Jampani},\ and\ \citenamefont
  {Lagerwall}}]{urbanski2017liquid}%
  \BibitemOpen
  \bibfield  {author} {\bibinfo {author} {\bibfnamefont {M.}~\bibnamefont
  {Urbanski}}, \bibinfo {author} {\bibfnamefont {C.~G.}\ \bibnamefont {Reyes}},
  \bibinfo {author} {\bibfnamefont {J.}~\bibnamefont {Noh}}, \bibinfo {author}
  {\bibfnamefont {A.}~\bibnamefont {Sharma}}, \bibinfo {author} {\bibfnamefont
  {Y.}~\bibnamefont {Geng}}, \bibinfo {author} {\bibfnamefont {V.~S.~R.}\
  \bibnamefont {Jampani}}, \ and\ \bibinfo {author} {\bibfnamefont {J.~P.}\
  \bibnamefont {Lagerwall}},\ }\href@noop {} {\bibfield  {journal} {\bibinfo
  {journal} {J. Phys.: Condens. Matter}\ }\textbf {\bibinfo {volume} {29}},\
  \bibinfo {pages} {133003} (\bibinfo {year} {2017})}\BibitemShut {NoStop}%
\bibitem [{\citenamefont {Nelson}\ and\ \citenamefont
  {Peliti}(1987)}]{nelson:fluctuations}%
  \BibitemOpen
  \bibfield  {author} {\bibinfo {author} {\bibfnamefont {D.~R.}\ \bibnamefont
  {Nelson}}\ and\ \bibinfo {author} {\bibfnamefont {L.}~\bibnamefont
  {Peliti}},\ }\href@noop {} {\bibfield  {journal} {\bibinfo  {journal} {J.
  Phys. France}\ }\textbf {\bibinfo {volume} {48}},\ \bibinfo {pages} {1085}
  (\bibinfo {year} {1987})}\BibitemShut {NoStop}%
\bibitem [{\citenamefont {Helfrich}\ and\ \citenamefont
  {Prost}(1988)}]{helfrich:intrinsic}%
  \BibitemOpen
  \bibfield  {author} {\bibinfo {author} {\bibfnamefont {W.}~\bibnamefont
  {Helfrich}}\ and\ \bibinfo {author} {\bibfnamefont {J.}~\bibnamefont
  {Prost}},\ }\href@noop {} {\bibfield  {journal} {\bibinfo  {journal} {Phys.
  Rev. A}\ }\textbf {\bibinfo {volume} {38}},\ \bibinfo {pages} {3065}
  (\bibinfo {year} {1988})}\BibitemShut {NoStop}%
\bibitem [{\citenamefont {Nguyen}\ \emph {et~al.}(2013)\citenamefont {Nguyen},
  \citenamefont {Geng}, \citenamefont {Selinger},\ and\ \citenamefont
  {Selinger}}]{nguyen:nematic}%
  \BibitemOpen
  \bibfield  {author} {\bibinfo {author} {\bibfnamefont {T.-S.}\ \bibnamefont
  {Nguyen}}, \bibinfo {author} {\bibfnamefont {J.}~\bibnamefont {Geng}},
  \bibinfo {author} {\bibfnamefont {R.~L.~B.}\ \bibnamefont {Selinger}}, \ and\
  \bibinfo {author} {\bibfnamefont {J.~V.}\ \bibnamefont {Selinger}},\
  }\href@noop {} {\bibfield  {journal} {\bibinfo  {journal} {Soft Matter}\
  }\textbf {\bibinfo {volume} {9}},\ \bibinfo {pages} {8314} (\bibinfo {year}
  {2013})}\BibitemShut {NoStop}%
\bibitem [{\citenamefont {Virga}(1994)}]{virga:variational}%
  \BibitemOpen
  \bibfield  {author} {\bibinfo {author} {\bibfnamefont {E.~G.}\ \bibnamefont
  {Virga}},\ }\href@noop {} {\emph {\bibinfo {title} {Variational Theories for
  Liquid Crystals}}}\ (\bibinfo  {publisher} {Chapman \& Hall},\ \bibinfo
  {address} {London},\ \bibinfo {year} {1994})\BibitemShut {NoStop}%
\bibitem [{\citenamefont {Cartan}(1925)}]{cartan:geometrie}%
  \BibitemOpen
  \bibfield  {author} {\bibinfo {author} {\bibfnamefont {E.}~\bibnamefont
  {Cartan}},\ }\href@noop {} {\emph {\bibinfo {title} {La G\'eom\'etrie des
  Espaces de {R}iemann}}},\ \bibinfo {series} {M\'emorial des Sciences
  Math\'ematiques}, Vol.~\bibinfo {volume} {9}\ (\bibinfo  {publisher}
  {Gauthier-Villars},\ \bibinfo {address} {Paris},\ \bibinfo {year}
  {1925})\BibitemShut {NoStop}%
\bibitem [{\citenamefont {Levi-Civita}(1917)}]{levi-civita:nozione}%
  \BibitemOpen
  \bibfield  {author} {\bibinfo {author} {\bibfnamefont {T.}~\bibnamefont
  {Levi-Civita}},\ }\href@noop {} {\bibfield  {journal} {\bibinfo  {journal}
  {Rend. Circ. Matem. Palermo}\ }\textbf {\bibinfo {volume} {42}},\ \bibinfo
  {pages} {173} (\bibinfo {year} {1917})}\BibitemShut {NoStop}%
\bibitem [{\citenamefont {Rosso}\ \emph {et~al.}(2012)\citenamefont {Rosso},
  \citenamefont {Virga},\ and\ \citenamefont {Kralj}}]{rosso2012parallel}%
  \BibitemOpen
  \bibfield  {author} {\bibinfo {author} {\bibfnamefont {R.}~\bibnamefont
  {Rosso}}, \bibinfo {author} {\bibfnamefont {E.~G.}\ \bibnamefont {Virga}}, \
  and\ \bibinfo {author} {\bibfnamefont {S.}~\bibnamefont {Kralj}},\
  }\href@noop {} {\bibfield  {journal} {\bibinfo  {journal} {Continuum Mech.
  Thermodyn.}\ }\textbf {\bibinfo {volume} {24}},\ \bibinfo {pages} {643}
  (\bibinfo {year} {2012})}\BibitemShut {NoStop}%
\bibitem [{\citenamefont {Persico}(1921)}]{persico:realizzazione}%
  \BibitemOpen
  \bibfield  {author} {\bibinfo {author} {\bibfnamefont {E.}~\bibnamefont
  {Persico}},\ }\href@noop {} {\bibfield  {journal} {\bibinfo  {journal} {Atti
  R. Acc. Linc. Rend. Cl. Scienze Mat. Fis. Nat.}\ }\textbf {\bibinfo {volume}
  {30(V)}},\ \bibinfo {pages} {127} (\bibinfo {year} {1921})}\BibitemShut
  {NoStop}%
\bibitem [{\citenamefont {Pfister}(2002)}]{pfister:spatial}%
  \BibitemOpen
  \bibfield  {author} {\bibinfo {author} {\bibfnamefont {F.}~\bibnamefont
  {Pfister}},\ }\href@noop {} {\bibfield  {journal} {\bibinfo  {journal} {Proc.
  Instn. Mech. Engrs.}\ }\textbf {\bibinfo {volume} {216C}},\ \bibinfo {pages}
  {33} (\bibinfo {year} {2002})}\BibitemShut {NoStop}%
\bibitem [{\citenamefont {Selinger}\ \emph {et~al.}(2011)\citenamefont
  {Selinger}, \citenamefont {Konya}, \citenamefont {Travesset},\ and\
  \citenamefont {Selinger}}]{selinger:monte}%
  \BibitemOpen
  \bibfield  {author} {\bibinfo {author} {\bibfnamefont {R.~L.~B.}\
  \bibnamefont {Selinger}}, \bibinfo {author} {\bibfnamefont {A.}~\bibnamefont
  {Konya}}, \bibinfo {author} {\bibfnamefont {A.}~\bibnamefont {Travesset}}, \
  and\ \bibinfo {author} {\bibfnamefont {J.~V.}\ \bibnamefont {Selinger}},\
  }\href@noop {} {\bibfield  {journal} {\bibinfo  {journal} {J. Phys. Chem. B}\
  }\textbf {\bibinfo {volume} {115}},\ \bibinfo {pages} {13989} (\bibinfo
  {year} {2011})}\BibitemShut {NoStop}%
\bibitem [{\citenamefont {Chen}\ and\ \citenamefont
  {Kamien}(2009)}]{chen:nematic}%
  \BibitemOpen
  \bibfield  {author} {\bibinfo {author} {\bibfnamefont {B.~G.}\ \bibnamefont
  {Chen}}\ and\ \bibinfo {author} {\bibfnamefont {R.~D.}\ \bibnamefont
  {Kamien}},\ }\href@noop {} {\bibfield  {journal} {\bibinfo  {journal} {Eur.
  Phys. J. E}\ }\textbf {\bibinfo {volume} {28}},\ \bibinfo {pages} {315}
  (\bibinfo {year} {2009})}\BibitemShut {NoStop}%
\bibitem [{\citenamefont {Napoli}\ and\ \citenamefont
  {Vergori}(2012{\natexlab{a}})}]{napoli2012surface}%
  \BibitemOpen
  \bibfield  {author} {\bibinfo {author} {\bibfnamefont {G.}~\bibnamefont
  {Napoli}}\ and\ \bibinfo {author} {\bibfnamefont {L.}~\bibnamefont
  {Vergori}},\ }\href@noop {} {\bibfield  {journal} {\bibinfo  {journal} {Phys.
  Rev. E}\ }\textbf {\bibinfo {volume} {85}},\ \bibinfo {pages} {061701}
  (\bibinfo {year} {2012}{\natexlab{a}})}\BibitemShut {NoStop}%
\bibitem [{\citenamefont {Ericksen}(1966)}]{ericksen:inequalities}%
  \BibitemOpen
  \bibfield  {author} {\bibinfo {author} {\bibfnamefont {J.~L.}\ \bibnamefont
  {Ericksen}},\ }\href@noop {} {\bibfield  {journal} {\bibinfo  {journal}
  {Phys. Fluids}\ }\textbf {\bibinfo {volume} {9}},\ \bibinfo {pages} {1205}
  (\bibinfo {year} {1966})}\BibitemShut {NoStop}%
\bibitem [{\citenamefont {Kamien}(2002)}]{kamien:geometry}%
  \BibitemOpen
  \bibfield  {author} {\bibinfo {author} {\bibfnamefont {R.~D.}\ \bibnamefont
  {Kamien}},\ }\href@noop {} {\bibfield  {journal} {\bibinfo  {journal} {Rev.
  Mod. Phys.}\ }\textbf {\bibinfo {volume} {74}},\ \bibinfo {pages} {953}
  (\bibinfo {year} {2002})}\BibitemShut {NoStop}%
\bibitem [{\citenamefont {Vitelli}\ and\ \citenamefont
  {Nelson}(2004)}]{vitelli:defect}%
  \BibitemOpen
  \bibfield  {author} {\bibinfo {author} {\bibfnamefont {V.}~\bibnamefont
  {Vitelli}}\ and\ \bibinfo {author} {\bibfnamefont {D.~R.}\ \bibnamefont
  {Nelson}},\ }\href@noop {} {\bibfield  {journal} {\bibinfo  {journal} {Phys.
  Rev. E}\ }\textbf {\bibinfo {volume} {70}},\ \bibinfo {pages} {051105}
  (\bibinfo {year} {2004})}\BibitemShut {NoStop}%
\bibitem [{\citenamefont {Segatti}\ \emph {et~al.}(2014)\citenamefont
  {Segatti}, \citenamefont {Snarski},\ and\ \citenamefont
  {Veneroni}}]{segatti:equilibrium}%
  \BibitemOpen
  \bibfield  {author} {\bibinfo {author} {\bibfnamefont {A.}~\bibnamefont
  {Segatti}}, \bibinfo {author} {\bibfnamefont {M.}~\bibnamefont {Snarski}}, \
  and\ \bibinfo {author} {\bibfnamefont {M.}~\bibnamefont {Veneroni}},\
  }\href@noop {} {\bibfield  {journal} {\bibinfo  {journal} {Phys. Rev. E}\
  }\textbf {\bibinfo {volume} {90}},\ \bibinfo {pages} {012501} (\bibinfo
  {year} {2014})}\BibitemShut {NoStop}%
\bibitem [{\citenamefont {Segatti}\ \emph {et~al.}(2016)\citenamefont
  {Segatti}, \citenamefont {Snarski},\ and\ \citenamefont
  {Veneroni}}]{segatti2016analysis}%
  \BibitemOpen
  \bibfield  {author} {\bibinfo {author} {\bibfnamefont {A.}~\bibnamefont
  {Segatti}}, \bibinfo {author} {\bibfnamefont {M.}~\bibnamefont {Snarski}}, \
  and\ \bibinfo {author} {\bibfnamefont {M.}~\bibnamefont {Veneroni}},\
  }\href@noop {} {\bibfield  {journal} {\bibinfo  {journal} {Math. Mod. Meth.
  Appl. Sci.}\ }\textbf {\bibinfo {volume} {26}},\ \bibinfo {pages} {1865}
  (\bibinfo {year} {2016})}\BibitemShut {NoStop}%
\bibitem [{\citenamefont {Napoli}\ and\ \citenamefont
  {Vergori}(2012{\natexlab{b}})}]{napoli2012extrinsic}%
  \BibitemOpen
  \bibfield  {author} {\bibinfo {author} {\bibfnamefont {G.}~\bibnamefont
  {Napoli}}\ and\ \bibinfo {author} {\bibfnamefont {L.}~\bibnamefont
  {Vergori}},\ }\href@noop {} {\bibfield  {journal} {\bibinfo  {journal} {Phys.
  Rev. Lett.}\ }\textbf {\bibinfo {volume} {108}},\ \bibinfo {pages} {207803}
  (\bibinfo {year} {2012}{\natexlab{b}})}\BibitemShut {NoStop}%
\bibitem [{\citenamefont {Spivak}(1999)}]{spivak:diffferential3}%
  \BibitemOpen
  \bibfield  {author} {\bibinfo {author} {\bibfnamefont {M.}~\bibnamefont
  {Spivak}},\ }\href@noop {} {\emph {\bibinfo {title} {A Comprehensive
  Introduction to Differential Geometry}}},\ \bibinfo {edition} {3rd}\ ed.,\
  Vol.~\bibinfo {volume} {3}\ (\bibinfo  {publisher} {Publish or Perish},\
  \bibinfo {address} {Houston},\ \bibinfo {year} {1999})\BibitemShut {NoStop}%
\bibitem [{\citenamefont {Lubensky}\ and\ \citenamefont
  {Prost}(1992)}]{lubensky:orientational}%
  \BibitemOpen
  \bibfield  {author} {\bibinfo {author} {\bibfnamefont {T.}~\bibnamefont
  {Lubensky}}\ and\ \bibinfo {author} {\bibfnamefont {J.}~\bibnamefont
  {Prost}},\ }\href@noop {} {\bibfield  {journal} {\bibinfo  {journal} {J.
  Phys. II France}\ }\textbf {\bibinfo {volume} {2}},\ \bibinfo {pages} {371}
  (\bibinfo {year} {1992})}\BibitemShut {NoStop}%
\bibitem [{\citenamefont {Straley}(1971)}]{straley:liquid}%
  \BibitemOpen
  \bibfield  {author} {\bibinfo {author} {\bibfnamefont {J.~P.}\ \bibnamefont
  {Straley}},\ }\href@noop {} {\bibfield  {journal} {\bibinfo  {journal} {Phys.
  Rev. A}\ }\textbf {\bibinfo {volume} {4}},\ \bibinfo {pages} {675} (\bibinfo
  {year} {1971})}\BibitemShut {NoStop}%
\bibitem [{\citenamefont {Kamien}\ \emph {et~al.}(2009)\citenamefont {Kamien},
  \citenamefont {Nelson}, \citenamefont {Santangelo},\ and\ \citenamefont
  {Vitelli}}]{kamien:extrinsic}%
  \BibitemOpen
  \bibfield  {author} {\bibinfo {author} {\bibfnamefont {R.~D.}\ \bibnamefont
  {Kamien}}, \bibinfo {author} {\bibfnamefont {D.~R.}\ \bibnamefont {Nelson}},
  \bibinfo {author} {\bibfnamefont {C.~D.}\ \bibnamefont {Santangelo}}, \ and\
  \bibinfo {author} {\bibfnamefont {V.}~\bibnamefont {Vitelli}},\ }\href@noop
  {} {\bibfield  {journal} {\bibinfo  {journal} {Phys. Rev. E}\ }\textbf
  {\bibinfo {volume} {80}},\ \bibinfo {pages} {051703} (\bibinfo {year}
  {2009})}\BibitemShut {NoStop}%
\bibitem [{\citenamefont {Mbanga}\ \emph {et~al.}(2012)\citenamefont {Mbanga},
  \citenamefont {Grason},\ and\ \citenamefont
  {Santangelo}}]{mbanga:frustrated}%
  \BibitemOpen
  \bibfield  {author} {\bibinfo {author} {\bibfnamefont {B.~L.}\ \bibnamefont
  {Mbanga}}, \bibinfo {author} {\bibfnamefont {G.~M.}\ \bibnamefont {Grason}},
  \ and\ \bibinfo {author} {\bibfnamefont {C.~D.}\ \bibnamefont {Santangelo}},\
  }\href@noop {} {\bibfield  {journal} {\bibinfo  {journal} {Phys. Rev. Lett.}\
  }\textbf {\bibinfo {volume} {108}},\ \bibinfo {pages} {017801} (\bibinfo
  {year} {2012})}\BibitemShut {NoStop}%
\bibitem [{\citenamefont {Biscari}\ and\ \citenamefont
  {Terentjev}(2006)}]{biscari:nematic}%
  \BibitemOpen
  \bibfield  {author} {\bibinfo {author} {\bibfnamefont {P.}~\bibnamefont
  {Biscari}}\ and\ \bibinfo {author} {\bibfnamefont {E.~M.}\ \bibnamefont
  {Terentjev}},\ }\href@noop {} {\bibfield  {journal} {\bibinfo  {journal}
  {Phys. Rev. E}\ }\textbf {\bibinfo {volume} {73}},\ \bibinfo {pages} {051706}
  (\bibinfo {year} {2006})}\BibitemShut {NoStop}%
\bibitem [{\citenamefont {Santangelo}\ \emph {et~al.}(2007)\citenamefont
  {Santangelo}, \citenamefont {Vitelli}, \citenamefont {Kamien},\ and\
  \citenamefont {Nelson}}]{santangelo:geometric}%
  \BibitemOpen
  \bibfield  {author} {\bibinfo {author} {\bibfnamefont {C.~D.}\ \bibnamefont
  {Santangelo}}, \bibinfo {author} {\bibfnamefont {V.}~\bibnamefont {Vitelli}},
  \bibinfo {author} {\bibfnamefont {R.~D.}\ \bibnamefont {Kamien}}, \ and\
  \bibinfo {author} {\bibfnamefont {D.~R.}\ \bibnamefont {Nelson}},\
  }\href@noop {} {\bibfield  {journal} {\bibinfo  {journal} {Phys. Rev. Lett.}\
  }\textbf {\bibinfo {volume} {99}},\ \bibinfo {pages} {017801} (\bibinfo
  {year} {2007})}\BibitemShut {NoStop}%
\bibitem [{\citenamefont {Jiang}\ \emph {et~al.}(2007)\citenamefont {Jiang},
  \citenamefont {Huber}, \citenamefont {Pelcovits},\ and\ \citenamefont
  {Powers}}]{jiang:vesicle}%
  \BibitemOpen
  \bibfield  {author} {\bibinfo {author} {\bibfnamefont {H.}~\bibnamefont
  {Jiang}}, \bibinfo {author} {\bibfnamefont {G.}~\bibnamefont {Huber}},
  \bibinfo {author} {\bibfnamefont {R.~A.}\ \bibnamefont {Pelcovits}}, \ and\
  \bibinfo {author} {\bibfnamefont {T.~R.}\ \bibnamefont {Powers}},\
  }\href@noop {} {\bibfield  {journal} {\bibinfo  {journal} {Phys. Rev. E}\
  }\textbf {\bibinfo {volume} {76}},\ \bibinfo {pages} {031908} (\bibinfo
  {year} {2007})}\BibitemShut {NoStop}%
\bibitem [{\citenamefont {Frank}\ and\ \citenamefont
  {Kardar}(2008)}]{frank:defects}%
  \BibitemOpen
  \bibfield  {author} {\bibinfo {author} {\bibfnamefont {J.~R.}\ \bibnamefont
  {Frank}}\ and\ \bibinfo {author} {\bibfnamefont {M.}~\bibnamefont {Kardar}},\
  }\href@noop {} {\bibfield  {journal} {\bibinfo  {journal} {Phys. Rev. E}\
  }\textbf {\bibinfo {volume} {77}},\ \bibinfo {pages} {041705} (\bibinfo
  {year} {2008})}\BibitemShut {NoStop}%
\bibitem [{\citenamefont {Giomi}(2012)}]{giomi:hyperbolic}%
  \BibitemOpen
  \bibfield  {author} {\bibinfo {author} {\bibfnamefont {L.}~\bibnamefont
  {Giomi}},\ }\href@noop {} {\bibfield  {journal} {\bibinfo  {journal} {Phys.
  Rev. Lett.}\ }\textbf {\bibinfo {volume} {109}},\ \bibinfo {pages} {136101}
  (\bibinfo {year} {2012})}\BibitemShut {NoStop}%
\bibitem [{\citenamefont {Bowick}\ \emph {et~al.}(2004)\citenamefont {Bowick},
  \citenamefont {Nelson},\ and\ \citenamefont
  {Travesset}}]{bowick:curvature-induced}%
  \BibitemOpen
  \bibfield  {author} {\bibinfo {author} {\bibfnamefont {M.}~\bibnamefont
  {Bowick}}, \bibinfo {author} {\bibfnamefont {D.~R.}\ \bibnamefont {Nelson}},
  \ and\ \bibinfo {author} {\bibfnamefont {A.}~\bibnamefont {Travesset}},\
  }\href@noop {} {\bibfield  {journal} {\bibinfo  {journal} {Phys. Rev. E}\
  }\textbf {\bibinfo {volume} {69}},\ \bibinfo {pages} {041102} (\bibinfo
  {year} {2004})}\BibitemShut {NoStop}%
\bibitem [{\citenamefont {Evans}(1995)}]{evans:phase}%
  \BibitemOpen
  \bibfield  {author} {\bibinfo {author} {\bibfnamefont {R.}~\bibnamefont
  {Evans}},\ }\href@noop {} {\bibfield  {journal} {\bibinfo  {journal} {J.
  Phys. II France}\ }\textbf {\bibinfo {volume} {5}},\ \bibinfo {pages} {507}
  (\bibinfo {year} {1995})}\BibitemShut {NoStop}%
\bibitem [{\citenamefont {Chen}(2016)}]{chen:theory}%
  \BibitemOpen
  \bibfield  {author} {\bibinfo {author} {\bibfnamefont {J.~Z.}\ \bibnamefont
  {Chen}},\ }\href@noop {} {\bibfield  {journal} {\bibinfo  {journal} {Prog.
  Polym. Sci.}\ }\textbf {\bibinfo {volume} {54--55}},\ \bibinfo {pages} {3}
  (\bibinfo {year} {2016})}\BibitemShut {NoStop}%
\bibitem [{\citenamefont {Vitelli}\ and\ \citenamefont
  {Turner}(2004)}]{vitelli:anomalous}%
  \BibitemOpen
  \bibfield  {author} {\bibinfo {author} {\bibfnamefont {V.}~\bibnamefont
  {Vitelli}}\ and\ \bibinfo {author} {\bibfnamefont {A.~M.}\ \bibnamefont
  {Turner}},\ }\href@noop {} {\bibfield  {journal} {\bibinfo  {journal} {Phys.
  Rev. Lett.}\ }\textbf {\bibinfo {volume} {93}},\ \bibinfo {pages} {215301}
  (\bibinfo {year} {2004})}\BibitemShut {NoStop}%
\bibitem [{\citenamefont {Bowick}\ and\ \citenamefont
  {Giomi}(2009)}]{bowick:two-dimensional}%
  \BibitemOpen
  \bibfield  {author} {\bibinfo {author} {\bibfnamefont {M.~J.}\ \bibnamefont
  {Bowick}}\ and\ \bibinfo {author} {\bibfnamefont {L.}~\bibnamefont {Giomi}},\
  }\href@noop {} {\bibfield  {journal} {\bibinfo  {journal} {Adv. Phys.}\
  }\textbf {\bibinfo {volume} {58}},\ \bibinfo {pages} {449} (\bibinfo {year}
  {2009})}\BibitemShut {NoStop}%
\bibitem [{\citenamefont {Abbena}\ \emph {et~al.}(2006)\citenamefont {Abbena},
  \citenamefont {Salamon},\ and\ \citenamefont {Gray}}]{gray:differential}%
  \BibitemOpen
  \bibfield  {author} {\bibinfo {author} {\bibfnamefont {E.}~\bibnamefont
  {Abbena}}, \bibinfo {author} {\bibfnamefont {S.}~\bibnamefont {Salamon}}, \
  and\ \bibinfo {author} {\bibfnamefont {A.}~\bibnamefont {Gray}},\ }\href@noop
  {} {\emph {\bibinfo {title} {Modern Differential Geometry of Curves and
  Surfaces with Mathematica}}},\ \bibinfo {edition} {3rd}\ ed.\ (\bibinfo
  {publisher} {Chapman and Hall/CRC},\ \bibinfo {address} {Boca Raton, FL},\
  \bibinfo {year} {2006})\BibitemShut {NoStop}%
\end{thebibliography}%
\bibliographystyle{apsrev4-1}

\end{document}